\documentclass[12pt]{article}
\usepackage{amsmath}
\usepackage{amssymb}
\usepackage{graphics}

\makeatletter\@addtoreset{equation}{section}
\makeatother

\allowdisplaybreaks[1]
\begin{document}
\begin{titlepage}
\begin{flushright}
TIT/HEP-621\\
September 2012
\end{flushright}
\vspace{0.5cm}
\begin{center}
{\Large \bf
Torsion and Supersymmetry in $\Omega$-background
}
\lineskip .75em
\vskip0.5cm
{\large Katsushi Ito${}^{1}$, Hiroaki Nakajima${}^{2}$ and Shin
Sasaki${}^{3}$ }
\vskip 2.5em
${}^{1}$ {\normalsize\it Department of Physics,\\
Tokyo Institute of Technology\\
Tokyo, 152-8551, Japan} \vskip 1.0em
${}^{2}$ {\normalsize\it Department of Physics 
and Center for Theoretical Sciences,\\
National Taiwan University\\
Taipei, 10617, Taiwan, R.O.C.} \vskip 1.0em
${}^{3}$ {\normalsize\it Department of Physics,\\
Kitasato University\\
Sagamihara, 252-0373, Japan}
\vskip 3.0em
\end{center}
\begin{abstract}
We study the dimensional reduction of ten-dimensional super Yang-Mills 
theory in curved backgrounds with torsion.
We examine the parallel spinor conditions and the constraints for the 
torsion parameters which preserve supersymmetry and gauge symmetry 
in four dimensions.
In particular we examine the ten-dimensional $\Omega$-background with 
the  torsion which is identified with the R-symmetry Wilson line gauge fields. 
After the dimensional reduction, we obtain the 
$\Omega$-deformed ${\cal N}=4$ super Yang-Mills theory. 
Solving the parallel spinor conditions and the torsion constraints, 
we classify the deformed supersymmetry associated with the 
topological twist of ${\cal N}=4$ supersymmetry.
We also study deformed supersymmetries in the Nekrasov-Shatashvili limit.
\end{abstract}
\end{titlepage}

\baselineskip=0.7cm
\section{Introduction}
The $\Omega$-background \cite{Moore:1997dj}
has been recognized as an interesting and useful deformation for studying 
non-perturbative effects in supersymmetric gauge theories 
via the localization technique 
\cite{Nekrasov:2002qd,Losev:2003py,Nekrasov:2003rj}. 
This background can be embedded into superstrings and 
the instanton partition functions are extracted from the scattering 
amplitudes \cite{Awata:2005fa, Iqbal:2007ii,
Huang:2010kf,Antoniadis:2010iq,Nakayama:2011be}. 
The microscopic deformed instanton effective action is also obtained from the 
D3/D($-1$) brane system in the R-R 3-form backgrounds 
\cite{Billo:2006jm,Ito:2010vx}. 

The $\Omega$-background is a curved geometry  with the action of $U(1)$ 
vector fields and is realized in higher dimensions. 
The background breaks the Poincar\'e symmetry and also
supersymmetry in general. 
A part of the supersymmetries, however, can be recovered by introducing 
the R-symmetry Wilson line gauge fields. 
For example, the $\Omega$ deformation of $\mathcal{N}=2$ super Yang-Mills 
theory is obtained by the dimensional reduction of six-dimensional 
$\mathcal{N}=1$ theory in the geometry with $U(1)^2$-action and the 
$SU(2)$ R-symmetry Wilson line gauge fields.
One can recover a scalar supersymmetry by choosing the 
appropriate Wilson lines, which is obtained by the topological twist of 
${\cal N}=2$ supersymmetry. 
Using this equivariant scalar supercharge, we can apply the localization 
method to compute the instanton partition function \cite{Nekrasov:2002qd}. 

In the previous paper \cite{Ito:2011cr}, we have studied 
${\cal N}=1$ super Yang-Mills theory in ten-dimensional 
$\Omega$-background with the $U(1)^6$-action and the constant 
$SU(4)$ R-symmetry Wilson line gauge fields. 
After the dimensional reduction to four dimensions, we have obtained 
the $\Omega$-deformed ${\cal N}=4$ super Yang-Mills theory. 
This theory admits the deformed supersymmetry in some cases. 
For the self-dual $\Omega$-background, the theory is invariant under the 
anti-chiral half of the ${\cal N}=4$ supersymmetry. 
For the ten-dimensional $\Omega$-background restricted to the 
six-dimensional $\Omega$-background with the appropriate 
Wilson line gauge fields, it corresponds to the ${\cal N}=2^*$ deformation 
of ${\cal N}=4$ theory. 
The explicit construction of the deformed supersymmetry transformations 
are however a very cumbersome task due to the complicated 
form of the deformed Lagrangian. 

The purpose of the present work is to study systematically the supersymmetry 
of four-dimensional $\Omega$-deformed ${\cal N}=4$ super Yang-Mills theory 
from the viewpoint of ten-dimensional ${\cal N}=1$ theory 
in a curved background.
For the flat spacetime background the dimensional reduction of ${\cal N}=1$ 
supersymmetry leads to the ${\cal N}=4$ supersymmetry in four dimensions 
\cite{Brink:1976bc}. 
The supersymmetry for the ${\cal N}=1$ super Yang-Mills theory 
in a curved manifold leads to the parallel spinor conditions, 
which implies that the curved background has the Ricci-flat special holonomy. 
This was generalized into the curved spacetime with the Killing spinor 
conditions \cite{Blau:2000xg}. 
Recently the localizations of ${\cal N}=4$ and ${\cal N}=2^*$ 
theories on the sphere have been studied with the help of the supersymmetry 
associated with the Killing spinor conditions 
\cite{Pestun:2007rz,Okuda:2010ke}.

For supersymmetry in the $\Omega$-background, it is necessary to 
introduce the R-symmetry Wilson line gauge fields, which is not 
realized by the deformation of the metric. 
In order to study this Wilson line deformation, we will investigate 
more general set-up, namely, 
${\cal N}=1$ super Yang-Mills theory in a curved 
background with torsion. 
The parallel spinor conditions are modified due to the torsion, which 
relaxes the Ricci-flat conditions for the curved spacetime. 
Note that generic torsion is inconsistent with gauge invariance. 
We will consider a special class of torsion such that the resulting 
four-dimensional gauge theory is gauge invariant after the dimensional 
reduction. 
When we apply this formulation to the ten-dimensional 
$\Omega$-background, we can identify the torsion with the R-symmetry 
Wilson line gauge fields. By solving the modified parallel spinor 
conditions we will find the constraints for the $\Omega$-background
parameters and the Wilson line gauge fields. 

As in the ${\cal N}=2$ case, 
the deformed scalar supersymmetries can be constructed 
by the topological twist of the ${\cal N}=4$ supersymmetry, 
which is classified as follows: 
the half-twist, the Vafa-Witten twist and the Marcus twist \cite{Ya}. 
We will construct the deformed supersymmetries for these twists.
We will further study the Nekrasov-Shatashvili limit \cite{Nekrasov:2009rc} 
of the deformed 
supersymmetry, where the supersymmetry is enhanced due to the partial recovery 
of the Poincar\'e invariance in two dimensions. 

This paper is organized as follows: 
in section 2, we introduce the ten-dimensional ${\cal N}=1$ super 
Yang-Mills theory in a curved background with torsion. 
In section 3, we study the dimensional reduction to four dimensions and 
examine the conditions such that the reduced theory has the gauge 
symmetry and also supersymmetry, which becomes the parallel spinor 
conditions and the constraints for the torsion. 
In section 4, we study the $\Omega$-background with torsion, which is 
identified with the Wilson line gauge fields. 
We examine 
the parallel spinor conditions and the torsion constraints and obtain 
the conditions for the deformation parameters. 
We then construct the deformed supersymmetries associated with the 
various twists as well as the Nekrasov-Shatashvili limit. 
In the appendix, we summarize the Dirac matrices in four and six dimensions.

\section{Ten-dimensional super Yang-Mills theory in curved background 
with torsion}
In this section, we introduce ten-dimensional $\mathcal{N}=1$ super Yang-Mills 
theory in curved background with torsion and discuss supersymmetry 
in the background. 

We first define ten-dimensional $\mathcal{N}=1$ super Yang-Mills theory 
with gauge group $G$ in the flat spacetime. 
This theory contains a gauge field $A_{M}$ $(M=0,1,\ldots,9)$ 
and a Majorana-Weyl fermion $\Psi$, where both fields 
belong to the adjoint representation of $G$. 
The Lagrangian is 
\begin{align}
\mathcal{L}_{0}=
\frac{1}{\kappa g^{2}}\mathrm{Tr}\biggl[
-\frac{1}{4}F^{MN}F_{MN}
-\frac{i}{2}\bar{\Psi}\Gamma^{M}D_{M}\Psi
\biggr],
\label{flatlag}
\end{align}
where $g$ is the coupling constant, 
$F_{MN}=\partial_{M}A_{N}-\partial_{N}A_{M}+i[A_{M},A_{N}]$ is 
the field strength of $A_{M}$. 
The gamma matrices $\Gamma^{M}$ are 
defined by $\Gamma^{M}\Gamma^{N}+\Gamma^{N}\Gamma^{M}=2\eta^{MN}$, 
where the flat metric $\eta_{MN}$ is taken to be Lorentzian as 
$\eta_{MN}=\mathrm{diag}(-1,+1,\ldots,+1)$. 
The gauge covariant derivative is defined by 
$D_{M}\ast=\partial_{M}\ast{}+i[A_{M},\ast]$. 
We normalize the generators $T^u$ $(u=1,\ldots,\mathrm{dim}G)$ 
of the gauge group $G$ 
as $\mathrm{Tr}(T^{u}T^{v})=\kappa\delta^{uv}$.

The Lagrangian \eqref{flatlag} is invariant up to a total derivative 
under the supersymmetry transformation \cite{Brink:1976bc}
\begin{gather}
\delta A_{M}=i\bar{\zeta}\,\Gamma_{M}\Psi,\quad 
\delta\Psi=-\frac{1}{2}F_{MN}\Gamma^{[M}\Gamma^{N]}\zeta, 
\label{transf0}
\end{gather}
where $\zeta$ is a constant Majorana-Weyl spinor. 
The square bracket in 
$\Gamma^{[M}\Gamma^{N]}$ denotes the antisymmetrization of the indices 
in the product of two gamma matrices, defined by 
$\Gamma^{[M}\Gamma^{N]}=
\frac{1}{2}(\Gamma^{M}\Gamma^{N}-\Gamma^{N}\Gamma^{M})$. 
$\Gamma^{[M_{1}}\Gamma^{M_{2}}\cdots \Gamma^{M_{n}]}$ is similarly 
normalized by the factor $1/n!$. 
The variation of the Lagrangian under \eqref{transf0} is 
\begin{align}
\delta\mathcal{L}_{0}
&=
\frac{1}{\kappa g^{2}}\mathrm{Tr}\biggl[
-\frac{1}{2}\bar{\Psi}\Gamma^{M}
[\bar{\Psi}\Gamma_{M}\zeta,\Psi]
+\frac{i}{2}\bar{\Psi}\Gamma^{[M}\Gamma^{N}\Gamma^{P]}\zeta\,(D_{[M}F_{NP]})
\notag\\
&\qquad\qquad{}
-\frac{i}{4}D_{M}(\bar{\Psi}\Gamma^{[N}\Gamma^{P]}\Gamma^{M}
\zeta\,F_{NP})
\biggr].
\label{variation1}
\end{align}
Since the first and the second terms
vanish by the Fierz and the Bianchi identity, 
respectively and the third term is a total derivative, 
the action is invariant under (\ref{transf0}). 

We next consider 
ten-dimensional curved spacetime background which is 
represented by the metric $G_{\mathcal{MN}}$ 
and the torsion $T_{\mathcal{MN}}{}^{\mathcal{P}}$ 
(see, for example \cite{Freedman:2012zz}). 
We use the calligraphic 
letters $\mathcal{M}$, $\mathcal{N}$, $\mathcal{P}$, \ldots\ 
$({}=0,1,\ldots,9)$ 
for the indices of the curved spacetime coordinates. 
We also introduce the vielbein $e_{\mathcal{M}}^{M}$, 
where the capital letters $M$, $N$, $P$, \ldots\ 
are used for the indices of the tangent space coordinates. 
The metric is written in terms of the vielbein as 
$G_{\mathcal{MN}}^{}=\eta_{MN}^{}e_{\mathcal{M}}^{M}e_{\mathcal{N}}^{N}$. 
The spin connection $\widehat{\omega}_{\mathcal{M},NP}$ is related to 
the vielbein and the torsion by Cartan's first structure equation 
\begin{align}
T_{\mathcal{MN}}{}^{P}
=
\partial_{\mathcal{M}}e_{\mathcal{N}}^{P}
-\partial_{\mathcal{N}}e_{\mathcal{M}}^{P}
+\widehat{\omega}_{\mathcal{M},}{}^{P}{}_{Q}\,e_{\mathcal{N}}^{Q}
-\widehat{\omega}_{\mathcal{N},}{}^{P}{}_{Q}\,e_{\mathcal{M}}^{Q}. 
\end{align}
Then $\widehat{\omega}_{\mathcal{M},NP}$ is expressed in terms of 
$e_{\mathcal{M}}^{M}$ and $T_{\mathcal{MN}}{}^{\mathcal{P}}$ as 
\begin{align}
\widehat{\omega}_{\mathcal{M},NP}
&=
\omega_{\mathcal{M},NP}+K_{\mathcal{M},NP},
\label{spcon}
\\
\omega_{\mathcal{M},NP}
&=
\frac{1}{2}\bigl(
C_{\mathcal{M}N,P}-C_{NP,\mathcal{M}}+C_{P\mathcal{M},N}
\bigr),
\\
K_{\mathcal{M},NP}
&=
-\frac{1}{2}\bigl(
T_{\mathcal{M}N,P}-T_{NP,\mathcal{M}}+T_{P\mathcal{M},N}
\bigr),
\label{contorsion}
\end{align}
where $C_{\mathcal{MN}}{}^{P}$ are the Ricci rotation coefficients defined by 
\begin{align}
C_{\mathcal{MN}}{}^{P}=
\partial_{\mathcal{M}}e_{\mathcal{N}}^{P}
-\partial_{\mathcal{N}}e_{\mathcal{M}}^{P}.
\label{ric}
\end{align}
The tensor $K_{\mathcal{M},NP}$ is called the contorsion. 
The torsion is expressed in terms of the contorsion as 
\begin{align}
T_{\mathcal{MN}}{}^{\mathcal{P}}
=
-K_{\mathcal{M},\,\mathcal{N}}{}^{\mathcal{P}}
+K_{\mathcal{N},\,\mathcal{M}}{}^{\mathcal{P}}.
\label{invert}
\end{align}
We also introduce the affine connection 
$\widehat{\varGamma}_{\mathcal{MN}}{}^{\mathcal{P}}$. 
Then the covariant derivative of the vielbein is 
\begin{align}
\widehat{\nabla}_{\mathcal{M}}e_{\mathcal{N}}^{P}
=
\partial_{\mathcal{M}}e_{\mathcal{N}}^{P}
-\widehat{\varGamma}_{\mathcal{MN}}{}^{\mathcal{P}}\,e_{\mathcal{P}}^{P}
+\widehat{\omega}_{\mathcal{M},}{}^{P}{}_{Q}\,e_{\mathcal{N}}^{Q}. 
\end{align}
Here $\widehat{\nabla}_{\mathcal{M}}$ denotes 
the spacetime covariant derivative including the torsion. 
We also denote the spacetime covariant derivative without the torsion 
as $\nabla_{\mathcal{M}}$. 
We relate the two connections by imposing the vielbein postulate
\begin{align}
\widehat{\nabla}_{\mathcal{M}}e_{\mathcal{N}}^{P}=0.
\label{pos}
\end{align}
{}From \eqref{spcon} and \eqref{pos}, 
$\widehat{\varGamma}_{\mathcal{MN}}{}^{\mathcal{P}}$ is decomposed into 
the torsion-independent part and the contorsion part as 
\begin{align}
\widehat{\varGamma}_{\mathcal{MN}}{}^{\mathcal{P}}
=
\varGamma_{\mathcal{MN}}{}^{\mathcal{P}}
+K_{\mathcal{M},}{}^{\mathcal{P}}{}_{\mathcal{N}}, 
\end{align}
where the first term is the usual Christoffel symbol 
(the Levi-Civita connection)
\begin{align}
\varGamma_{\mathcal{MN}}{}^{\mathcal{P}}
=
\frac{1}{2}G^{\mathcal{PQ}}
(\partial_{\mathcal{M}}G_{\mathcal{NQ}}+\partial_{\mathcal{N}}G_{\mathcal{MQ}}
-\partial_{\mathcal{Q}}G_{\mathcal{MN}}). 
\label{chr2}
\end{align}

Now we introduce the Lagrangian 
in the curved spacetime background with the torsion 
by replacing all the derivative $\partial_{\mathcal{M}}$ in \eqref{flatlag} 
to the spacetime covariant derivative $\widehat{\nabla}_{\mathcal{M}}$ 
and the appropriate contraction of the indices. 
For the vector field and the spinor field, $\widehat{\nabla}_{\mathcal{M}}$ 
acts as 
\begin{align}
\widehat{\nabla}_{\mathcal{M}}A_{\mathcal{N}}
=
\partial_{\mathcal{M}}A_{\mathcal{N}}
-\widehat{\varGamma}_{\mathcal{MN}}{}^{\mathcal{P}}A_{\mathcal{P}},
\quad 
\widehat{\nabla}_{\mathcal{M}}\Psi
=
\biggl(\partial_{\mathcal{M}}+\frac{1}{2}\widehat{\omega}_{\mathcal{M},NP}
\Gamma^{NP}\biggr)\Psi,
\label{eq:covariant_derivative}
\end{align}
where $\Gamma^{MN}=\frac{1}{2}\Gamma^{[M}\Gamma^{N]}$ is the ten-dimensional 
Lorentz generator. 
The field strength $F_{\mathcal{M}\mathcal{N}}$ is replaced 
with $\widehat{F}_{\mathcal{M}\mathcal{N}}$ defined by 
\begin{align}
\widehat{F}_{\mathcal{M}\mathcal{N}}
&=
\widehat{\nabla}_{\mathcal{M}}A_{\mathcal{N}}
-\widehat{\nabla}_{\mathcal{N}}A_{\mathcal{M}}
+i[A_{\mathcal{M}},A_{\mathcal{N}}]
\notag\\
&=
F_{\mathcal{M}\mathcal{N}}
-T_{\mathcal{M}\mathcal{N}}{}^{\mathcal{P}}A_{\mathcal{P}}. 
\label{modF}
\end{align}
The gauge covariant derivative $D_{\mathcal{M}}$ is replaced with 
$\widehat{\nabla}_{\mathcal{M}}^{(G)}\ast=\widehat{\nabla}_{\mathcal{M}}\ast
{}+i[A_{\mathcal{M}},\ast]$ 
which is covariant with respect to both the gauge 
and the general coordinate transformation. 
Then the Lagrangian in the curved background with the torsion 
becomes 
\begin{align}
\widehat{\mathcal{L}}
&=
\frac{1}{\kappa g^{2}}\mathrm{Tr}\biggl[
-\frac{1}{4}e\bigl(e^{\mathcal{M}}_{M}e^{\mathcal{N}}_{N}
\widehat{F}_{\mathcal{M}\mathcal{N}}\bigr)^{2}
-\frac{i}{2}e\,\bar{\Psi}e^{\mathcal{M}}_{M}\Gamma^{M}
\widehat{\nabla}_{\mathcal{M}}^{(G)}\Psi
\biggr],
\label{curvedlag}
\end{align}
where $e$ is the determinant of the vielbein and $e^{\mathcal{M}}_{M}$ 
is the inverse vielbein. 
We note that 
\eqref{curvedlag} is not gauge invariant when the torsion is nonzero, since 
the last term in \eqref{modF} explicitly depends on the gauge field itself. 

We discuss the invariance of the action under 
the supersymmetry transformation: 
\begin{align}
\delta A_{\mathcal{M}}
=
ie_{\mathcal{M}}^{M}\,\bar{\zeta}\,\Gamma_{M}\Psi, \quad
\delta\Psi
=
-\frac{1}{2}e^{\mathcal{M}}_{M}e^{\mathcal{N}}_{N}
\widehat{F}_{\mathcal{M}\mathcal{N}}\Gamma^{[M}\Gamma^{N]}\zeta. 
\label{SUSYtransf}
\end{align}
The variation of the Lagrangian \eqref{curvedlag} becomes 
\begin{align}
\delta\widehat{\mathcal{L}}
&=
\frac{1}{\kappa g^{2}}\mathrm{Tr}\biggl[
\frac{i}{2}e\bar{\Psi}
\Gamma^{[\mathcal{M}}\Gamma^{\mathcal{N}}\Gamma^{\mathcal{P}]}
\zeta\bigl(
\widehat{\nabla}_{\mathcal{[M}}^{(G)}\widehat{F}_{\mathcal{N}\mathcal{P}]}^{}
\bigr)
-\frac{i}{4}e\widehat{\nabla}_{\mathcal{M}}
(\bar{\Psi}\Gamma^{[\mathcal{N}}\Gamma^{\mathcal{P}]}\Gamma^{\mathcal{M}}
\zeta\,\widehat{F}_{\mathcal{N}\mathcal{P}})
\notag\\
&\qquad\qquad{}
+\frac{i}{2}e\bar{\Psi}\Gamma^{\mathcal{M}}
\Gamma^{[\mathcal{N}}\Gamma^{\mathcal{P}]}
\widehat{F}_{\mathcal{N}\mathcal{P}}(\widehat{\nabla}_{\mathcal{M}}\zeta)
\biggr],
\label{variation}
\end{align}
where we have used the Fierz identity. 
In the first term we will compute 
$\widehat{\nabla}_{[\mathcal{M}}^{(G)}
\widehat{F}_{\mathcal{N}\mathcal{P}]}^{}$. 
{}From the Bianchi identity, we obtain 
\begin{align}
\widehat{\nabla}_{[\mathcal{M}}^{(G)}\widehat{F}_{\mathcal{N}\mathcal{P}]}^{}
=
-(\partial_{\mathcal{Q}}A_{[\mathcal{M}})
T_{\mathcal{N}\mathcal{P}]}{}^{\mathcal{Q}}
-(\partial_{[\mathcal{M}}T_{\mathcal{N}\mathcal{P}]}{}^{\mathcal{R}}
+T_{[\mathcal{M}\mathcal{N}}{}^{\mathcal{Q}}
T_{\mathcal{P}]\mathcal{Q}}{}^{\mathcal{R}})A_{\mathcal{R}}.
\label{precon}
\end{align}
In order that the first term of \eqref{precon} vanishes, 
the torsion must be zero. 
In this case 
the second term in \eqref{variation} becomes a total derivative. 
The last term in \eqref{variation} becomes zero 
by requiring that $\zeta$ satisfies the parallel spinor condition 
\begin{align}
\nabla_{\mathcal{M}}\zeta=0. 
\label{eq:parallel_spinor0}
\end{align}
Hence the Lagrangian \eqref{curvedlag} is neither invariant under 
the supersymmetry transformation \eqref{SUSYtransf} 
nor gauge invariant in ten dimensions, 
unless the torsion vanishes. 
It is related to the fact that bosonic and fermionic 
physical degrees of freedom 
are different since the gauge field becomes massive. 
This implies that the action is not supersymmetric. 
However, if we consider the dimensional reduction, 
the action becomes invariant under the gauge and supersymmetry 
transformations when the torsion satisfies certain conditions, 
as we will see in the next section. 

\section{Dimensional reduction and parallel spinor conditions}
We now consider the dimensional reduction of the theory \eqref{curvedlag} 
to four dimensions. We also perform the Wick rotation $x^{0}=-ix^{10}$.
The local Lorentz group $SO(10)$ is reduced to 
$SU(2)_{L}\times SU(2)_{R}\times SU(4)$, 
where $SU(2)_L\times SU(2)_R$ is the Lorentz group in four dimensions and 
$SU(4)$ becomes the R-symmetry of the reduced theory. 
After the dimensional reduction, the gauge field $A_{\mathcal{M}}$ 
is decomposed as $A_{\mathcal{M}}=(A_{\mu},\varphi_{\mathcal{A}}^{})$, 
where $A_{\mu}$ $(\mu=1,\ldots,4)$ is the gauge field and 
$\varphi_{\mathcal{A}}^{}$ $(\mathcal{A}=5,\ldots,10)$ are the scalar fields 
in four dimensions. 
The spinor field $\Psi$ is also decomposed as 
$\Psi=(\Lambda_{\alpha}^{A},\bar{\Lambda}^{\dot{\alpha}}_{A})$, 
where $\alpha,\dot{\alpha}=1,2$ are the $SU(2)_{L}$ and $SU(2)_{R}$ indices 
respectively. 
These indices are raised and lowered by the antisymmetric $\varepsilon$-symbol 
normalized as $\varepsilon^{12}=-\varepsilon_{12}=1$. 
$A=1,\ldots,4$ is the index for the vector representation of $SU(4)$. 
The gamma matrices are decomposed as 
\begin{align}
\Gamma^{M}=
\left(
-i\begin{pmatrix} 0 & \sigma^{m}_{\alpha\dot{\alpha}} \\ 
\bar{\sigma}^{m\dot{\alpha}\alpha} & 0 \end{pmatrix}
\otimes \boldsymbol{1}_{8},\ 
\begin{pmatrix} \boldsymbol{1}_{2} & 0 \\ 0 & -\boldsymbol{1}_{2} \end{pmatrix}
\otimes 
\begin{pmatrix} 0 & \Sigma^{aAB} \\ \bar{\Sigma}^{a}_{AB} & 0 \end{pmatrix}
\right),
\end{align}
where we have decomposed the index $M$ as 
$M=(m,a)$ $(m=1,\ldots,4,\ a=5,\ldots,10)$. $\boldsymbol{1}_{n}$ denotes the 
$n\times n$ identity matrix. The conventions of four- and six-dimensional 
Dirac matrices $\sigma^{m},\bar{\sigma}^{m},\Sigma^{a},\bar{\Sigma}^{a}$ 
are given in the appendix. 

The fields and the background do not depend on the internal coordinates 
$x^{\mathcal{A}}$ in the dimensional reduction. 
By setting $\partial_{\mathcal{A}}=0$ in \eqref{curvedlag}, 
we obtain the four-dimensional Lagrangian as 
\begin{align}
\widehat{\mathcal{L}}_{\mathrm{4D}}
&=
\frac{1}{\kappa g^{2}}\mathrm{Tr}\biggl[
\frac{1}{4}e
\Bigl(e^{\mu}_{m}e^{\nu}_{n}
\widehat{F}_{\mu\nu}
+(e^{\mu}_{m}e^{\mathcal{A}}_{n}-e^{\mu}_{n}e^{\mathcal{A}}_{m})
\widehat{F}_{\mu\mathcal{A}}
+e^{\mathcal{A}}_{m}e^{\mathcal{B}}_{n}
\widehat{F}_{\mathcal{AB}}
\Bigr)^{2}
\notag\\
&\qquad\qquad\ {}
+\frac{1}{2}e
\Bigl(e^{\mu}_{m}e^{\nu}_{a}
\widehat{F}_{\mu\nu}
+(e^{\mu}_{m}e^{\mathcal{A}}_{a}-e^{\mu}_{a}e^{\mathcal{A}}_{m})
\widehat{F}_{\mu\mathcal{A}}
+e^{\mathcal{A}}_{m}e^{\mathcal{B}}_{a}
\widehat{F}_{\mathcal{AB}}
\Bigr)^{2}
\notag\\
&\qquad\qquad\ {}
+\frac{1}{4}e
\Bigl(e^{\mu}_{a}e^{\nu}_{b}
\widehat{F}_{\mu\nu}
+(e^{\mu}_{a}e^{\mathcal{A}}_{b}-e^{\mu}_{b}e^{\mathcal{A}}_{a})
\widehat{F}_{\mu\mathcal{A}}
+e^{\mathcal{A}}_{a}e^{\mathcal{B}}_{b}
\widehat{F}_{\mathcal{AB}}
\Bigr)^{2}
\notag\\
&\qquad\qquad\ {}
+e_{m}^{\mu}\Lambda^{\alpha A}\sigma^{m}_{\alpha\dot{\alpha}}
D_{\mu}\bar{\Lambda}^{\dot{\alpha}}_{A}
+ie_{m}^{\mathcal{A}}\Lambda^{\alpha A}\sigma^{m}_{\alpha\dot{\alpha}}
[\varphi_{\mathcal{A}},\bar{\Lambda}^{\dot{\alpha}}_{A}]
\notag\\
&\qquad\qquad\ {}
-\frac{1}{2}e_{a}^{\mathcal{A}}\bar{\Sigma}^{a}_{AB}
\Lambda^{\alpha A}[\varphi_{\mathcal{A}},\Lambda^{B}_{\alpha}]
-\frac{1}{2}e_{a}^{\mathcal{A}}\Sigma^{aAB}
\bar{\Lambda}_{\dot{\alpha}A}[\varphi_{\mathcal{A}},\Lambda_{B}^{\dot{\alpha}}]
\notag\\
&\qquad\qquad\ {}
+\frac{i}{2}e_{a}^{\mu}\bar{\Sigma}^{a}_{AB}
\Lambda^{\alpha A}D_{\mu}\Lambda^{B}_{\alpha}
+\frac{i}{2}e_{a}^{\mu}\Sigma^{aAB}
\bar{\Lambda}_{\dot{\alpha}A}D_{\mu}\bar{\Lambda}_{B}^{\dot{\alpha}}
+\frac{1}{4}\widehat{\omega}_{m,np}
\Lambda^{A}\epsilon^{mnpq}\sigma_{q}\bar{\Lambda}_{A}
\notag\\
&\qquad\qquad\ {}
+\frac{i}{4}(\widehat{\omega}_{a,mn}+2\widehat{\omega}_{m,na})
(\Lambda^{A}\sigma^{mn}\bar{\Sigma}^{a}_{AB}\Lambda^{B}
+\bar{\Lambda}_{A}\bar{\sigma}^{mn}\Sigma^{aAB}\bar{\Lambda}_{B})
\notag\\
&\qquad\qquad\ {}
+\frac{1}{2}(2\widehat{\omega}_{a,mb}-\widehat{\omega}_{m,ab})
\Lambda^{A}\sigma^{m}(\bar{\Sigma}^{ab})_{A}{}^{B}\bar{\Lambda}_{B}
\notag\\
&\qquad\qquad\ {}
-\frac{i}{8}\widehat{\omega}_{a,bc}
(\Lambda^{A}(\bar{\Sigma}^{[a}\Sigma^{b}\bar{\Sigma}^{c]})_{AB}\Lambda^{B}
+\bar{\Lambda}_{A}(\Sigma^{[a}\bar{\Sigma}^{b}\Sigma^{c]})^{AB}\Lambda_{B})
\biggr],
\label{4Dlag}
\end{align}
where 
$\epsilon^{mnpq}$ is the totally antisymmetric tensor normalized as 
$\epsilon^{1234}=1$. 
$\sigma^{mn}$, $\bar{\sigma}^{mn}$ and $\Sigma^{ab}$, $\bar{\Sigma}^{ab}$ 
are the Lorentz generators in four and six dimensions respectively, 
which are defined in the appendix. 
$\widehat{F}_{\mu\nu}$, $\widehat{F}_{\mu\mathcal{A}}$, 
$\widehat{F}_{\mathcal{AB}}$ are 
the components of the modified field strength \eqref{modF}, 
which are obtained as 
\begin{align}
\widehat{F}_{\mu\nu}
&=
F_{\mu\nu}-T_{\mu\nu}{}^{\rho}A_{\rho}
-T_{\mu\nu}{}^{\mathcal{A}}\varphi_{\mathcal{A}},
\notag\\
\widehat{F}_{\mu\mathcal{A}}
&=
D_{\mu}\varphi_{\mathcal{A}}-T_{\mu\mathcal{A}}{}^{\rho}A_{\rho}
-T_{\mu\mathcal{A}}{}^{\mathcal{B}}\varphi_{\mathcal{B}},
\notag\\
\widehat{F}_{\mathcal{AB}}
&=
i[\varphi_{\mathcal{A}},\varphi_{\mathcal{B}}]
-T_{\mathcal{A}\mathcal{B}}{}^{\rho}A_{\rho}
-T_{\mathcal{A}\mathcal{B}}{}^{\mathcal{C}}\varphi_{\mathcal{C}}.
\end{align}
The Lagrangian \eqref{4Dlag} does not have the gauge invariance since 
$\widehat{\mathcal{L}}_{\mathrm{4D}}$ depends on $A_{\mu}$ explicitly. 
However the gauge invariance is recovered by setting 
\begin{align}
T^{}_{\mathcal{MN}}{}^{\rho}=
(T^{}_{\mathcal{\mu\nu}}{}^{\rho},\ T^{}_{\mu\mathcal{A}}{}^{\rho},\ 
T^{}_{\mathcal{AB}}{}^{\rho})=0. 
\label{condition1}
\end{align}

Next 
we examine the supersymmetry in the dimensionally 
reduced theory under the gauge invariance condition \eqref{condition1}. 
We use the notation in ten dimensions for convenience. 
The condition for supersymmetry is that \eqref{variation} becomes 
a total derivative. 
\eqref{precon} vanishes due to \eqref{condition1} and the reduction 
$\partial_{\mathcal{A}}=0$. 
The second and the third terms become zero by imposing the condition 
\begin{align}
\partial_{[\mathcal{M}}T_{\mathcal{N}\mathcal{P}]}{}^{\mathcal{R}}
+T_{[\mathcal{M}\mathcal{N}}{}^{\mathcal{Q}}
T_{\mathcal{P}]\mathcal{Q}}{}^{\mathcal{R}}=0.
\label{condition4}
\end{align}
The second term in the variation \eqref{variation} does not become 
a total derivative. We have 
\begin{align}
e\,\widehat{\nabla}_{\mathcal{M}}V^{\mathcal{M}}=
\partial_{\mathcal{M}}(e\,V^{\mathcal{M}})
+e\,T_{\mathcal{M}\mathcal{N}}{}^{\mathcal{M}}V^{\mathcal{N}}, 
\label{formula}
\end{align}
where the vector $V^{\mathcal{M}}$ is given by
\begin{align}
V^{\mathcal{M}}
=
\frac{1}{\kappa g^{2}}\mathrm{Tr}\biggl[
-\frac{i}{4}\bar{\Psi}\Gamma^{\mathcal{N}\mathcal{P}}\Gamma^{\mathcal{M}}
\zeta\,\widehat{F}_{\mathcal{N}\mathcal{P}}
\biggr]. 
\end{align}
Hence we have the traceless condition 
such that the second term in \eqref{formula} vanishes as 
\begin{align}
T_{\mathcal{M}\mathcal{N}}{}^{\mathcal{M}}=0.
\label{condition3}
\end{align}
The last term in \eqref{variation} becomes zero
when $\zeta$ satisfies the parallel spinor condition 
modified by the torsion as 
\begin{align}
\widehat{\nabla}_{\mathcal{M}}\zeta=0. 
\label{eq:parallel_spinor}
\end{align}
Therefore the dimensionally reduced theory from \eqref{curvedlag} is invariant 
under the supersymmetry transformation \eqref{SUSYtransf} generated by 
the parallel spinor $\zeta$ satisfying \eqref{eq:parallel_spinor} 
when the torsion $T_{\mathcal{MN}}{}^{\mathcal{P}}$ satisfies 
the conditions \eqref{condition1}, \eqref{condition4} and \eqref{condition3}. 
After the dimensional reduction, 
the supersymmetry transformation \eqref{SUSYtransf}
becomes 
\begin{align}
\delta A_{\mu}
&=
-e_{\mu}^{m}\zeta^{A}\sigma_{m}\bar{\Lambda}_{A}
-e_{\mu}^{m}\bar{\zeta}^{A}\bar{\sigma}_{m}\Lambda_{A}
+ie_{\mu}^{\mathcal{A}}\zeta^{A}\bar{\Sigma}_{aAB}\Lambda^{B}
-ie_{\mu}^{\mathcal{A}}\bar{\zeta}_{A}\Sigma^{AB}_{a}\bar{\Lambda}_{B},
\notag\\
\delta \Lambda^{A}
&=
\sigma^{mn}\zeta^{A}\Bigl(e^{\mu}_{m}e^{\nu}_{a}
\widehat{F}_{\mu\nu}
+(e^{\mu}_{m}e^{\mathcal{A}}_{n}-e^{\mu}_{n}e^{\mathcal{A}}_{m})
\widehat{F}_{\mu\mathcal{A}}
+e^{\mathcal{A}}_{m}e^{\mathcal{B}}_{n}
\widehat{F}_{\mathcal{AB}}
\Bigr)
\notag\\
&\quad{}
+i\Sigma^{aAB}\sigma^{m}\bar{\zeta}_{B}
\Bigl(e^{\mu}_{m}e^{\nu}_{a}
\widehat{F}_{\mu\nu}
+(e^{\mu}_{m}e^{\mathcal{A}}_{a}-e^{\mu}_{a}e^{\mathcal{A}}_{m})
\widehat{F}_{\mu\mathcal{A}}
+e^{\mathcal{A}}_{m}e^{\mathcal{B}}_{a}
\widehat{F}_{\mathcal{AB}}
\Bigr)
\notag\\
&\quad{}
+(\Sigma^{ab})^{A}{}_{B}\zeta^{B}
\Bigl(e^{\mu}_{a}e^{\nu}_{b}
\widehat{F}_{\mu\nu}
+(e^{\mu}_{a}e^{\mathcal{A}}_{b}-e^{\mu}_{b}e^{\mathcal{A}}_{a})
\widehat{F}_{\mu\mathcal{A}}
+e^{\mathcal{A}}_{a}e^{\mathcal{B}}_{b}
\widehat{F}_{\mathcal{AB}}
\Bigr),
\notag\\
\delta \bar{\Lambda}_{A}
&=
\bar{\sigma}^{mn}\bar{\zeta}_{A}\Bigl(e^{\mu}_{m}e^{\nu}_{a}
\widehat{F}_{\mu\nu}
+(e^{\mu}_{m}e^{\mathcal{A}}_{n}-e^{\mu}_{n}e^{\mathcal{A}}_{m})
\widehat{F}_{\mu\mathcal{A}}
+e^{\mathcal{A}}_{m}e^{\mathcal{B}}_{n}
\widehat{F}_{\mathcal{AB}}
\Bigr)
\notag\\
&\quad{}
+i\bar{\Sigma}^{a}_{AB}\bar{\sigma}^{m}\zeta^{B}
\Bigl(e^{\mu}_{m}e^{\nu}_{a}
\widehat{F}_{\mu\nu}
+(e^{\mu}_{m}e^{\mathcal{A}}_{a}-e^{\mu}_{a}e^{\mathcal{A}}_{m})
\widehat{F}_{\mu\mathcal{A}}
+e^{\mathcal{A}}_{m}e^{\mathcal{B}}_{a}
\widehat{F}_{\mathcal{AB}}
\Bigr)
\notag\\
&\quad{}
+(\bar{\Sigma}^{ab})_{A}{}^{B}\bar{\zeta}_{B}
\Bigl(e^{\mu}_{a}e^{\nu}_{b}
\widehat{F}_{\mu\nu}
+(e^{\mu}_{a}e^{\mathcal{A}}_{b}-e^{\mu}_{b}e^{\mathcal{A}}_{a})
\widehat{F}_{\mu\mathcal{A}}
+e^{\mathcal{A}}_{a}e^{\mathcal{B}}_{b}
\widehat{F}_{\mathcal{AB}}
\Bigr),
\notag\\
\delta \varphi_{\mathcal{A}}
&=
-e^{m}_{\mathcal{A}}\zeta^{A}\sigma_{m}\bar{\Lambda}_{A}
-e^{m}_{\mathcal{A}}\bar{\zeta}^{A}\bar{\sigma}_{m}\Lambda_{A}
+ie^{a}_{\mathcal{A}}\zeta^{A}\bar{\Sigma}_{aAB}\Lambda^{B}
-ie^{a}_{\mathcal{A}}\bar{\zeta}_{A}\Sigma^{AB}_{a}\bar{\Lambda}_{B},
\label{susy_transf2}
\end{align}
where we decomposed $\zeta$ as 
$\zeta=(\zeta^{A}_{\alpha},\bar{\zeta}_{\dot{\alpha}A})$.

Here we consider the case of the flat spacetime and that 
the constant torsion is turned only in the six-dimensional direction as
an example. 
In the Lagrangian \eqref{4Dlag}, the torsion gives the mass terms for 
the fermions and scalars. The supersymmetry conditions restrict the form 
of the mass terms. From \eqref{condition4} 
the constant torsion satisfies $T_{[ab}{}^{d}T_{c]d}{}^{e}=0$, 
which implies that the torsion can be regarded as the structure constant of 
a Lie algebra $\mathcal{T}$ and forms the adjoint representation of 
$\mathcal{T}$. The dimension of $\mathcal{T}$ is equal to or less than six, 
where the latter case is possible when 
the components of the torsion are not linearly independent. 
The traceless condition 
$T_{ab}{}^{a}=0$ from \eqref{condition3} must be also satisfied. 

If the torsion is totally antisymmetric, the traceless condition 
is satisfied. 
Moreover, from \eqref{contorsion} 
the contorsion is proportional to the torsion as 
$K_{a,bc}=-\frac{1}{2}T_{ab,c}$. In this case 
$\mathcal{T}$ becomes a subalgebra of $SU(4)$. 
The parallel spinor condition \eqref{eq:parallel_spinor} becomes 
\begin{align}
T_{ab,c}(\Sigma^{bc})^{A}{}_{B}\zeta^{B}_{\alpha}=0,\quad 
T_{ab,c}(\bar{\Sigma}^{bc})_{A}{}^{B}\bar{\zeta}_{\dot{\alpha}B}=0.
\label{ps2}
\end{align}
Since the matrices acting on the parallel spinors in \eqref{ps2} are 
the Hermitian conjugate to each other, 
we have the same 
number of the left-handed and the right-handed parallel spinors. 
The number of supersymmetries
depends on the choice of $\mathcal{T}$ and how $\mathcal{T}$ is embedded 
into $SU(4)$. 
We summarize the relation between $\mathcal{T}$
and the number of supersymmetry in table~\ref{tb:torsion}. 
\begin{table}[t]
\begin{center}
\begin{tabular}{|l|c|}
\hline
$\mathcal{T}$ & $\sharp$ of SUSY \\
\hline 
$SU(2)\times SU(2)$, $SU(2)\times U(1)^{2}$, $U(1)^{3}$ & $\mathcal{N}=0$ \\
\hline
$SU(2)\times U(1)$, $U(1)^{2}$ & $\mathcal{N}=1$ \\
\hline
$SU(2)$, $U(1)$ & $\mathcal{N}=2$ \\
\hline
\end{tabular}
\caption{Torsion algebra $\mathcal{T}\in SU(4)$ and 
the number $\mathcal{N}$ of supersymmetry. The 
embedding of $\mathcal{T}$ is chosen such that $\mathcal{N}$ becomes maximal. }
\label{tb:torsion}
\end{center}
\end{table}

As an example, we consider the case $\mathcal{T}=U(1)^{2}$. 
If $\mathcal{T}$ is embedded into $SU(4)$ appropriately, 
the parallel spinor condition \eqref{ps2} has the form 
\begin{align}
\begin{pmatrix} 0 & 0 & 0 & 0 \\ 0 & t_{1a} & 0 & 0 \\
0 & 0 & t_{2a} & 0 \\ 0 & 0 & 0 & t_{3a} \end{pmatrix}^{\!\!A}_{\ B} 
\zeta^{B}=0, \quad 
t_{1a}+t_{2a}+t_{3a}=0,\quad 
t_{1a},t_{2a},t_{3a}\neq 0. 
\label{ps3}
\end{align}
The similar condition for $\bar{\zeta}_{A}$ holds. 
The solution to \eqref{ps3} is 
$\zeta^{A}=(\zeta^{1},0,0,0)^{T}$. Then we have $\mathcal{N}=1$ supersymmetry. 
We can check that it corresponds to the $\mathcal{N}=1^{*}$ deformation 
\cite{DoHoKhMa}. 
The mass term for the fermions takes the form of 
\begin{align}
\mathcal{L}_{\mathrm{m}}=
m_{AB}\Lambda^{A}\Lambda^{B}+\bar{m}^{AB}\bar{\Lambda}_{A}\bar{\Lambda}_{B},
\label{massterm}
\end{align}
where the two mass matrices $m_{AB}$ and $\bar{m}^{AB}$ are defined by 
\begin{align}
m_{AB}=\frac{i}{16}
(\bar{\Sigma}^{[a}\Sigma^{b}\bar{\Sigma}^{c]})_{AB}T_{ab,c},
\quad 
\bar{m}^{AB}=\frac{i}{16}
(\Sigma^{[a}\bar{\Sigma}^{b}\Sigma^{c]})^{AB}T_{ab,c}. 
\end{align}
When the parallel spinor condition becomes the form of \eqref{ps3}, 
we can show that 
each mass matrix has one zero eigenvalue. 
This is the $\mathcal{N}=1^{*}$ deformation. 
Similarly, in the case of $\mathcal{T}=U(1)$ we obtain the 
$\mathcal{N}=2^{*}$ deformation.

\section{$\Omega$-background and Deformed Supersymmetry}
In this section, we study the four-dimensional 
$\mathcal{N} = 4$ super Yang-Mills theory in the $\Omega$-background
with torsion. We solve the parallel spinor and the torsion conditions 
obtained in the previous section and classify the supersymmetries.

\subsection{$\mathcal{N} = 4$ super Yang-Mills theory in $\Omega$-background}
The Lagrangian 
of the four-dimensional $\mathcal{N} = 4$ super Yang-Mills theory in the
$\Omega$-background is obtained by the dimensional reduction of the
ten-dimensional $\mathcal{N} = 1$ super Yang-Mills theory in the 
spacetime with the metric:
\begin{eqnarray}
\begin{aligned}
 & d s^2 = (d x^m + \Omega^m {}_a d x^{a})^2 + (dx^{a})^2, 
\\
 & \Omega^m {}_a = \Omega^{mn} {}_a x_n, \quad \Omega^{mn} {}_a = -
 \Omega^{nm} {}_a, 
\end{aligned}
\label{eq:Omega_bg}
\end{eqnarray}
where $x^m$ and $x^a$ are the coordinates of the four- and six-dimensional 
spaces. 
The anti-symmetric matrices $\Omega_{mna}$ are parameterized as 
\begin{eqnarray}
\Omega_{mna} = 
\left(
\begin{array}{cccc}
0 & \epsilon_{1a} & 0 & 0 \\
- \epsilon_{1a} & 0 & 0 & 0 \\
0 & 0 & 0 & - \epsilon_{2a} \\
0 & 0 & \epsilon_{2a} & 0 
\end{array}
\right),
\end{eqnarray}
where $\epsilon_{1a}, \epsilon_{2a}$ are real parameters. 
These matrices commute with each other 
\begin{equation}
\Omega_{m} {}^p {}_a \Omega_{pnb} - \Omega_m {}^p {}_b \Omega_{pna} = 0.
\end{equation}
The vielbein is given by 
\begin{eqnarray}
\begin{aligned}
 & e^M {}_{\mathcal{M}} = 
\left(
\begin{array}{cc}
e^m {}_{\mu} & e^a {}_{\mu} \\
e^m {}_{\mathcal{A}} & e^a {}_{\mathcal{A}}
\end{array}
\right)
= 
\left(
\begin{array}{cc}
\delta^m {}_{\mu} & 0 \\
\delta_{\mathcal{A}}^a \Omega^m {}_{a} & \delta^a {}_{\mathcal{A}}
\end{array}
\right), \\
 & 
e^{\mathcal{M}} {}_M = 
\left(
\begin{array}{cc}
e^{\mu} {}_m & e^{\mathcal{A}} {}_m \\
e^{\mu} {}_a & e^{\mathcal{A}} {}_a
\end{array}
\right)
= 
\left(
\begin{array}{cc}
\delta^{\mu} {}_m & 0 \\
- \delta^{\mu}_m \Omega^{m} {}_a & \delta^{\mathcal{A}} {}_a
\end{array}
\right).
\end{aligned}
\label{vielbein}
\end{eqnarray}

We introduce the constant torsion along the internal directions,  
which is consistent with the dimensional reduction.
We want to study the supersymmetry conditions in this setup.
Another way to recover parts of supersymmetry is to introduce the constant
Wilson line gauge field $\mathcal{A}_a$ by gauging the 
$SU(4)$ R-symmetry \cite{Nekrasov:2003rj, Ito:2011cr}.
From the expression of the covariant derivative 
\eqref{eq:covariant_derivative} in the general curved background, 
we find that the $SU(4)$ R-symmetry Wilson line gauge field is identified with the
 contorsion through the following relation,
\begin{equation}
K_{\mathcal{A},bc} = - i \delta_{\mathcal{A}}^a 
(\mathcal{A}_a)^A {}_B (\Sigma_{bc})^B {}_A,
\label{eq:WL-contorsion}
\end{equation}
or equivalently
\begin{equation}
(\mathcal{A}_a)^A {}_B = \frac{i}{4} \delta^{\mathcal{A}}_a 
(\Sigma_{bc})^A {}_B K_{\mathcal{A},bc}.
\label{eq:Contorsion-WL}
\end{equation}
The other components except $K_{\mathcal{A},bc}$ are zero. 
{}From \eqref{invert}, the non-trivial components of the torsion are given by 
\begin{eqnarray}
T^{}_{\mathcal{AB}}{}^{\mu} &=&
-\delta^{a}_{\mathcal{A}}\delta^{b}_{\mathcal{B}}
(K_{a,b}{}^{c}-K_{b,a}{}^{c})\Omega^{\mu}_{c},
\label{unwanted} \\
T^{}_{\mathcal{AB}}{}^{\mathcal{C}}
&=&-\delta^{a}_{\mathcal{A}}\delta^{b}_{\mathcal{B}}\delta_{c}^{\mathcal{C}}
(K_{a,b}{}^{c}-K_{b,a}{}^{c}). 
\label{remaining}
\end{eqnarray}
The non-zero components of the spin and affine connections are evaluated as 
\begin{eqnarray}
\widehat{\omega}_{\mathcal{A}, mn} &=& \delta_{\mathcal{A}}^a \Omega_{mna}, 
\quad \widehat{\omega}_{\mathcal{A},bc} = K_{\mathcal{A},bc}, \\
\widehat{\varGamma}_{\mu \mathcal{A}} {}^{\nu} &=& \Omega^{\nu} {}_{\mu \mathcal{A}}, \\
\widehat{\varGamma}_{\mathcal{A} \mathcal{B}} {}^{\mu} &=& \Omega^{\mu}{}_{\rho
 \mathcal{A}} \Omega^{\rho} {}_{\mathcal{B}} 
- \delta_{\mathcal{B}}^b \Omega^{\mu}{}_{c} 
K_{\mathcal{A},}{}^{c}{}_{b}, \\
\widehat{\varGamma}_{\mathcal{A} \mathcal{B}} {}^{\mathcal{C}} &=& 
 \delta_{\mathcal{B}}^b \delta^{\mathcal{C}}_c K_{\mathcal{A},}{}^{c}{}_{b}.
\end{eqnarray}
The gauge invariance condition \eqref{condition1} reads 
\begin{equation}
T_{\mathcal{A} \mathcal{B}} {}^{\mu} = 0.
\end{equation}
Using this condition and substituting the vielbein \eqref{vielbein} and the torsion
\eqref{unwanted}, \eqref{remaining} into the Lagrangian \eqref{4Dlag}, 
we obtain
\begin{align}
 \mathcal{L}_{(\Omega,\mathcal{A})} = \frac{1}{\kappa g^2} \mathrm{Tr} 
\Big[ 
 &\, \frac{1}{4} F^{mn} F_{mn} 
+ \Lambda^A \sigma^{m} D_{m} \bar{\Lambda}_A + \frac{1}{2} \big( D_{m} \varphi_a -  F_{mn} \Omega^{n}_a \big)^2 \notag \\
 &\, - \frac{1}{2} (\Sigma_a )^{AB} \bar{\Lambda}_A [ \varphi_a , \bar{\Lambda}_B ] - \frac{1}{2} (\bar{\Sigma}_a )_{AB} \Lambda^A [ \varphi_a , \Lambda^B ] \notag \\
 &\, - \frac{1}{4} \Big( [ \varphi_a , \varphi_b ] + i \Omega^{m}_a D_{m} \varphi_b - i \Omega^{m}_b D_{m} \varphi_a - i F_{mn} \Omega^{m}_a \Omega^{n}_b \notag \\
 &\, \qquad \qquad - \frac{1}{2} \big( (\Sigma_b \bar{\Sigma}_c )^{A}{}_{B} \varphi_{c} (\mathcal{A}_a )^{B}{}_{A} - (\Sigma_a \bar{\Sigma}_c )^{A}{}_{B} \varphi_{c} (\mathcal{A}_b )^{B}{}_{A} \big) \Big)^2 \notag \\
 &\, - \frac{i}{2} \Omega^{m}_a \big( ( \Sigma_a )^{AB} \bar{\Lambda}_A D_{m} \bar{\Lambda}_B + (\bar{\Sigma}_a )_{AB} \Lambda^A D_{m} \Lambda^B \big) \notag \\
 &\, + \frac{i}{4} \Omega_{mn a} \big( (\Sigma_a )^{AB} \bar{\Lambda}_A \bar{\sigma}^{mn} \bar{\Lambda}_B + (\bar{\Sigma}_a )_{AB} \Lambda^A \sigma^{mn} \Lambda^B \big) \notag \\
 &\, + \frac{1}{2} (\Sigma_a )^{AB} \bar{\Lambda}_A \bar{\Lambda}_{D}
 (\mathcal{A}_a )^{D}{}_{B} - \frac{1}{2} (\bar{\Sigma}_a )_{AB}
 \Lambda^A (\mathcal{A}_a )^{B}{}_{D} \Lambda^{D} \Big].
\label{eq:N4SYMOmega}
\end{align}
This Lagrangian indeed coincides with the $\Omega$-deformed one with the
R-symmetry Wilson line obtained in \cite{Ito:2011cr}.

\subsection{Supersymmetry conditions in $\Omega$-background}
Now we examine the supersymmetry and gauge invariance conditions for the 
Lagrangian \eqref{eq:N4SYMOmega}.
We first write down the parallel spinor
condition \eqref{eq:parallel_spinor} in the $\Omega$-background with the torsion.
Then, we consider the constraints on the torsion 
\eqref{condition1}, \eqref{condition4} and \eqref{condition3}.

\subsubsection{Parallel spinor condition}
The parallel spinor condition \eqref{eq:parallel_spinor} in the
$\Omega$-background with the torsion is given by 
\begin{eqnarray}
& & \widehat{\nabla}_{\mu} \zeta = \partial_{\mu} \zeta = 0, 
\label{parallel4}
\\
& & \widehat{\nabla}_{\mathcal{A}} \zeta = 
\frac{1}{4} \delta^a {}_{\mathcal{A}} 
(\Omega_{mna} \Gamma^{mn} + K_{a,bc} \Gamma^{bc}) \zeta = 0.
\label{eq:spinor_condition}
\end{eqnarray}
From the condition \eqref{parallel4}, the parameter $\zeta$ becomes
constant.
When the $\Omega$-background matrices $\Omega_{mna}$ are
anti-self-dual or self-dual, and the torsion is zero, 
the condition \eqref{eq:spinor_condition} is satisfied 
for the chiral or anti-chiral parameters 
$\zeta^{A}_{\alpha}$, $\bar{\zeta}^{\dot{\alpha}}_A$ respectively.
In these cases, all the torsion conditions
\eqref{condition1}, \eqref{condition3} and
\eqref{condition4} are trivially satisfied and half of the $\mathcal{N} =
4$ supersymmetries 
are preserved \cite{Ito:2011cr}.
However, when $\Omega_{mna}$ is not (anti-)self-dual, the condition 
\eqref{eq:spinor_condition} can not be satisfied in general.

In following, we consider $\Omega_{mna}$ which is not (anti-)self-dual.
Since $\Gamma^{mn}$ and $\Gamma^{ab}$ are generators of four- and
six-dimensional rotations, $\Omega_{mna}$ and $K_{a,bc}$ are rotational
parameters of $SO(4)$ and $SO(6)$.
Since the 
matrices $\Omega_{mna}$ commute with each other, 
they are the rotational parameters of the $U(1)_L \times U(1)_R$ Cartan 
subgroup of the four-dimensional Lorentz group $SO(4) \simeq SU(2)_L
\times SU(2)_R$.
For the six-dimensional rotation group, 
we consider the subgroup $SO(2)' \times SO(4)' \simeq U(1)' \times SU(2)_{L'} \times
 SU(2)_{R'} $ of $SO(6)$.
We decompose the six-dimensional vector index $a$ 
into $a = (a', \hat{a})$ $(a'=5,6, \hat{a} = 
 7,8,9,10)$, associated with the $SO(2)'$ and $SO(4)'$ rotations
 respectively.
We cancel parts of the component in 
$\Omega_{mna} \Gamma^{mn}$ and $K_{a,bc} \Gamma^{bc}$ 
by identifying $U(1)$ charges of the Lorentz group with 
that of six dimensions. 
This is done by identifying
$SU(2)$'s in the Lorentz group with those in the R-symmetry
group.
These identifications 
correspond to the topological twist of the four-dimensional
$\mathcal{N} = 4$ supersymmetry \cite{Ya}.
Then the components of the contorsion $K_{a,bc}$ are the parameters of 
the Cartan subgroup $U(1)_{L'} \times U(1)_{R'}$ of 
$SU(2)_{L'} \times SU(2)_{R'}$, 
where $K_{a, \hat{b} \hat{c}}$ are non-zero and 
\begin{eqnarray}
K_{a, b'c'} = K_{a,b'\hat{c}} = 0.
\label{contorsion_zero}
\end{eqnarray}
The condition \eqref{eq:spinor_condition} becomes 
\begin{equation}
(\Omega_{mna} \Gamma^{mn} + K_{a, \hat{b} \hat{c}} \Gamma^{\hat{b}
 \hat{c}}) \zeta = 0.
\label{parallel_spinor_omega}
\end{equation}

Now we determine the $U(1)$ charges of the spinor parameter $\zeta =
(\zeta^A_{\alpha}, \bar{\zeta}^{\dot{\alpha}}_A)$. 
The representations $\mathbf{4}$ and $\bar{\mathbf{4}}$ of $SU(4)$ 
are decomposed into the representation of 
$U(1)' \times SU(2)_{L'} \times SU(2)_{R'}$:
\begin{eqnarray}
\mathbf{4} = (\mathbf{2}, \mathbf{1})_{1/2} + (\mathbf{1},
 \mathbf{2})_{-1/2}, \quad 
\bar{\mathbf{4}} = (\mathbf{2}, \mathbf{1})_{-1/2} + (\mathbf{1},
 \mathbf{2})_{1/2}, 
\end{eqnarray}
where the first and the second components in the parenthesis are 
the representations of $SU(2)_{L'}$ and $SU(2)_{R'}$ respectively.
The subscript $\pm 1/2$ denotes the $U(1)'$ charge and 
$\mathbf{2}$, $\mathbf{1}$ are the two-dimensional and the trivial
representation of each $SU(2)$.
Then we have the following decomposition of spinors,
\begin{equation}
\zeta^A_{\alpha} = (\zeta^{A'}_{\alpha}, \zeta^{\hat{A}}_{\alpha}),
 \quad 
\bar{\zeta}^{\dot{\alpha}}_{A} = (\bar{\zeta}^{\dot{\alpha}}_{A'},
\bar{\zeta}^{\dot{\alpha}}_{\hat{A}}), \quad A' = 1,2, \ \hat{A} = 3,4.
\end{equation}
Spinors that have $A' = 1,2$ are $\mathbf{2}$ representation of
$SU(2)_{R'}$ while that have $\hat{A} = 3,4$ are $\mathbf{2}$
representation of $SU(2)_{L'}$. 
The generators of 
$SO(4)'$ for the spinors $\zeta^{A'}_{\alpha}, 
\bar{\zeta}^{\dot{\alpha}}_{A'}$ and $\zeta_{\hat{A}}^{\alpha},
\bar{\zeta}_{\hat{A}}^{\dot{\alpha}} $ are $(\Sigma_{ab})_{A'} {}^{B'} = -
(\bar{\Sigma}_{ab})^{B'} {}_{A'}$ and $(\Sigma_{ab})_{\hat{A}} {}^{\hat{B}} = -
(\bar{\Sigma}_{ab})^{\hat{B}} {}_{\hat{A}}$.
We are interested in the $U(1)_X \ (X=L,R,L',R')$ charges of spinors associated with the Cartan
subgroups of $SU(2)_X$. 
We summarize the $U(1)$ charges for each spinor in table \ref{tb:U1_charge}.
\begin{table}[t]
\begin{center}
\begin{tabular}{lcccc}
\hline
 & $U(1)_L$ & $U(1)_R$ & $U(1)_{L'}$ & $U(1)_{R'}$ \\
\hline
$\zeta^{A'}_{\alpha}$ & $\pm \frac{1}{2}$ & $0$ & $0$ & $\pm \frac{1}{2}$ \\
$\zeta^{\hat{A}}_{\alpha}$ & $\pm \frac{1}{2}$ & $0$ & $\pm \frac{1}{2}$ & 0 \\
$\bar{\zeta}^{\dot{\alpha}}_{A'}$ & $0$ & $\pm \frac{1}{2}$ & $0$ &
				 $\pm \frac{1}{2}$ \\
$\bar{\zeta}^{\dot{\alpha}}_{\hat{A}}$ & $0$ & $\pm \frac{1}{2}$ &
			 $\pm \frac{1}{2}$ & $0$ \\
\hline
\end{tabular}
\caption{$U(1)_X$ charges of spinors.}
\label{tb:U1_charge}
\end{center}
\end{table}

There are three topological twists called the half twist \cite{Ya}, 
the Vafa-Witten twist \cite{Vafa:1994tf}
and the Marcus twist \cite{Marcus:1995mq}. 
\paragraph{Half twist}
In the half twist, $SU(2)_{R'}$ and $SU(2)_{R}$ are identified while the
$SU(2)_{L'}$ and $SU(2)_L$ are left intact. 
The new Lorentz group is defined as $SU(2)_L
\times [SU(2)_{R'} \times SU(2)_R]_{\mathrm{diag}}$ 
where $[SU(2)_{R'} \times SU(2)_R]_{\mathrm{diag}}$ denotes 
the diagonal subgroup of $SU(2)_{R'} \times SU(2)_{R}$.
Spinors $\zeta_{\alpha}^{A'}$, $\bar{\zeta}^{\dot{\alpha}}_{A'}$ 
can be decomposed under the new Lorentz group as 
 \begin{equation}
\zeta_{\alpha} {}^{A'} = (\sigma^m)_{\alpha B'} \epsilon^{A'B'} \zeta_m,
 \qquad 
\bar{\zeta}^{\dot{\alpha}} {}_{A'} = \delta^{\dot{\alpha}} {}_{A'}
\bar{\zeta} + (\bar{\sigma}^{mn})^{\dot{\alpha}} {}_{A'} \bar{\zeta}_{mn},
\label{eq:ht_parameters}
\end{equation}
where $\zeta_m$, $\bar{\zeta}$ and $\bar{\zeta}_{mn}$ are 
vector, scalar and anti-self-dual tensor respectively.

\paragraph{Vafa-Witten twist}
In the Vafa-Witten twist, $SU(2)_{L'} \times SU(2)_{R'}$ in the
R-symmetry and $SU(2)_R$ in the Lorentz symmetry is identified. 
The new Lorentz group is defined as
$SU(2)_L \times [SU(2)_{L'} \times SU(2)_{R'} \times
SU(2)_R]_{\mathrm{diag}}$. 
Spinors are decomposed as 
\begin{eqnarray}
& & \zeta_{\alpha} {}^{A'} = (\sigma^m)_{\alpha B'} \epsilon^{A'B'}
 \zeta_m, \qquad 
\bar{\zeta}^{\dot{\alpha}} {}_{A'} = \delta^{\dot{\alpha}} {}_{A'}
\bar{\zeta} + (\bar{\sigma}^{mn})^{\dot{\alpha}} {}_{A'}
\bar{\zeta}_{mn},
\label{eq:vwt_parameters1}
\\
& & \zeta_{\alpha} {}^{\hat{A}} = (\sigma^m)_{\alpha \hat{B}}
\epsilon^{\hat{A} \hat{B}}
 \zeta^{\prime}_m, \qquad 
\bar{\zeta}^{\dot{\alpha}} {}_{\hat{A}} = \delta^{\dot{\alpha}} {}_{\hat{A}}
\bar{\zeta}^{\prime} + (\bar{\sigma}^{mn})^{\dot{\alpha}} {}_{\hat{A}}
\bar{\zeta}^{\prime}_{mn}.
\label{eq:vwt_parameters2}
\end{eqnarray}

\paragraph{Marcus twist}
In the Marcus twist, 
$SU(2)_{L'}$ and $SU(2)_{L}$, $SU(2)_{R'}$ and $SU(2)_{R}$ are
identified.
The new Lorentz group is defined as 
$[SU(2)_{L'} \times SU(2)_{L}]_{\mathrm{diag}} 
\times [SU(2)_{R'} \times SU(2)_{R}]_{\mathrm{diag}}$.
Spinors are decomposed as 
\begin{eqnarray}
& & \zeta_{\alpha} {}^{A'} = (\sigma^m)_{\alpha B'} \epsilon^{A'B'}
 \zeta_m, \qquad 
\bar{\zeta}^{\dot{\alpha}} {}_{A'} = \delta^{\dot{\alpha}} {}_{A'}
\bar{\zeta} + (\bar{\sigma}^{mn})^{\dot{\alpha}} {}_{A'}
\bar{\zeta}_{mn},
\label{eq:mt_parameters1}
\\
& & \zeta_{\alpha} {}^{\hat{A}} = \delta_{\alpha} {}^{\hat{A}} \zeta +
 (\sigma^{mn})_{\alpha} {}^{\hat{A}} \zeta_{mn}, \qquad
 \bar{\zeta}^{\dot{\alpha}} {}_{\hat{A}} =
 (\bar{\sigma}^m)^{\dot{\alpha} \hat{B}} \epsilon_{\hat{A} \hat{B}} \bar{\zeta}_m.
\label{eq:mt_parameters2}
\end{eqnarray}

\subsubsection{Torsion conditions}
We have obtained the parallel spinor condition 
\eqref{parallel_spinor_omega}. 
Now we write down the conditions on the torsion 
\eqref{condition1}, \eqref{condition4} and \eqref{condition3} 
in the $\Omega$-background.
The gauge invariance condition \eqref{condition1} reads 
\begin{align}
(K_{a,b}{}^{c}-K_{b,a}{}^{c})\Omega_{mnc} = 0. 
\label{eq:torsion2}
\end{align}
The condition \eqref{condition4} becomes 
\begin{align}
T_{[ab} {}^d T_{c] d} {}^e = 0,
\label{eq:torsion3}
\end{align}
while the condition \eqref{condition3} is reduced to 
\begin{align}
T_{ab} {}^a = - K_{a,b} {}^a = 0.
\label{eq:torsion4}
\end{align}

We first consider the condition \eqref{eq:torsion2}. 
From \eqref{contorsion_zero}, the condition \eqref{eq:torsion2} becomes 
\begin{align}
K_{a',\hat{b}} {}^{\hat{c}} \Omega_{mn\hat{c}} &= 0, 
\label{eq:torsion11}
\\
(K_{\hat{a}, \hat{b}} {}^{\hat{c}} - K_{\hat{b}, \hat{a}}
 {}^{\hat{c}}) \Omega_{mn\hat{c}} &= 0.
\label{eq:torsion12}
\end{align}
Since these conditions are independent of $\Omega_{mna'}$, 
we assume that the parameters $\epsilon_{1a'},\epsilon_{2a'}$ are 
non-zero without loss of generality.

Next, the condition \eqref{eq:torsion3} becomes  
\begin{align}
T_{a'b'} {}^{\hat{d}} T_{\hat{c} \hat{d}} {}^{\hat{e}} + T_{b' \hat{c}}
 {}^{\hat{d}} T_{a' \hat{d}} {}^{\hat{e}} + T_{\hat{c} a'} {}^{\hat{d}}
 T_{b' \hat{d}} {}^{\hat{e}} &= 0, 
\label{eq:torsion31}
\\
T_{a' \hat{b}} {}^{\hat{d}} T_{\hat{c} \hat{d}} {}^{\hat{e}} +
 T_{\hat{b} \hat{c}} {}^{\hat{d}} T_{a' \hat{d}} {}^{\hat{e}} +
 T_{\hat{c} a'} {}^{\hat{d}} T_{\hat{b} \hat{d}} {}^{\hat{e}} &= 0, 
\label{eq:torsion32}
\\
T_{\hat{a}\hat{b}}{}^{\hat{d}}T_{\hat{c}\hat{d}}{}^{\hat{e}}
+T_{\hat{b}\hat{c}}{}^{\hat{d}}T_{\hat{a}\hat{d}}{}^{\hat{e}}
+T_{\hat{c}\hat{a}}{}^{\hat{d}}T_{\hat{b}\hat{d}}{}^{\hat{e}}
&= 0.
\label{eq:torsion33}
\end{align}
Using \eqref{invert}, the conditions \eqref{eq:torsion31} and 
\eqref{eq:torsion32} are rewritten as 
\begin{eqnarray}
 & &K_{b',\hat{c}} {}^{\hat{d}} K_{a',\hat{d}} {}^{\hat{e}} - K_{a',
 \hat{c}} {}^{\hat{d}} K_{b', \hat{d}} {}^{\hat{e}} =0, 
\label{eq:contorsion_commutation}
 \\
 & &K_{a', \hat{b}} {}^{\hat{d}} T_{\hat{c} \hat{d}} {}^{\hat{e}} + K_{a',
 \hat{c}} {}^{\hat{d}} T_{\hat{b} \hat{d}} {}^{\hat{e}} - K_{a'}
 {}^{\hat{e}} {}_{\hat{d}} T_{\hat{b} \hat{c}} {}^{\hat{d}} = 0.
\label{eq:U1_charge_condition}
\end{eqnarray}
The first equation \eqref{eq:contorsion_commutation} 
 is the commuting condition of the contorsion matrices $K_{a',
\hat{b}} {}^{\hat{c}}$.
Then $K_{a', \hat{b} \hat{c}}$ are parameters of $U(1)_{L'} \times U(1)_{R'}$.
Finally, the condition \eqref{eq:torsion4} becomes 
\begin{equation}
K_{\hat{a},\hat{b}} {}^{\hat{a}} = 0.
\label{eq:torsion41}
\end{equation}
In the following we solve the parallel spinor condition and the
 conditions on the torsion in each twist separately.

\subsection{Solutions to the conditions}
\subsubsection{Half twist \label{half_twist}}

First we consider the half twist. 
The supercharges $Q_{\alpha}^{A'}$ and $\bar{Q}^{\dot{\alpha}}_{A'}$ are 
decomposed into $\bar{Q}$, $Q_m$, $\bar{Q}_{mn}$. 
The parameters $\zeta_{\alpha}^{A'}$, $\bar{\zeta}^{\dot{\alpha}}_{A'}$ 
are decomposed in 
the same way. 
The parallel spinor condition \eqref{parallel_spinor_omega} for 
$\bar{\zeta}$, $\bar{\zeta}_{mn}$ and $\zeta_m$ are 
\begin{eqnarray}
& & \left[
\delta^{\dot{\beta}} {}_{A'}
\Omega_{mna} (\bar{\sigma}^{mn})^{\dot{\alpha}} {}_{\dot{\beta}}
+ 
\delta^{\dot{\alpha}} {}_{B'} K_{a,\hat{b}
 \hat{c}} (\bar{\Sigma}^{\hat{b} \hat{c}})_{A'} {}^{B'}
\right]
\bar{\zeta} = 0, 
\label{eq:ht_scalar_1}
\\
& & 
\left[
\delta^{\dot{\gamma}} {}_{A'}
\Omega_{mna} (\bar{\sigma}^{mn})^{\dot{\alpha}} {}_{\dot{\beta}}
(\bar{\sigma}^{pq})^{\dot{\beta}} {}_{\dot{\gamma}} +
\delta^{\dot{\beta}} {}_{B'} K_{a,\hat{b} \hat{c}}
(\bar{\Sigma}^{\hat{b} \hat{c}})_{A'} {}^{B'} (\bar{\sigma}^{pq})^{\dot{\alpha}} {}_{\dot{\beta}}
\right] \bar{\zeta}_{pq} = 0, 
\label{eq:ht_tensor}
\\
& & \left[
\delta^{\dot{\gamma}} {}_{C'}
\Omega_{mna} (\sigma^{mn})_{\alpha} {}^{\beta} (\sigma^p)_{\beta \dot{\gamma}}
\epsilon^{A'C'} + 
\delta^{\dot{\beta}} {}_{C'}
K_{a,\hat{b} \hat{c}} (\sigma^p)_{\alpha \dot{\beta}} \epsilon^{B'C'}
(\Sigma^{\hat{b} \hat{c}})^{A'} {}_{B'} 
\right] \zeta_p = 0.
\label{eq:ht_vector}
\end{eqnarray}
In order that the scalar supersymmetry is preserved, the $U(1)_{R}$ and
$U(1)_{R'}$ charges must be identified as 
\begin{equation}
\delta^{\dot{\alpha}} {}_{B'} K_{a,\hat{b}
 \hat{c}} (\bar{\Sigma}^{\hat{b} \hat{c}})_{A'} {}^{B'} = - 
\delta^{\dot{\beta}} {}_{A'}
\Omega_{mna} (\bar{\sigma}^{mn})^{\dot{\alpha}} {}_{\dot{\beta}}.
\label{eq:half_scalar}
\end{equation}
We then find that the 
scalar and one component of the tensor supersymmetries are preserved
and the others are broken under the condition
\eqref{eq:half_scalar}\footnote{We will discuss the tensor
supersymmetries in section 4.4}. 

We then solve the constraints on the torsion.
The anti-self-dual part of $K_{a, \hat{b} \hat{c}}$ and $\Omega_{mna}$ 
 are identified by the relation \eqref{eq:half_scalar}. 
Then the anti-self-dual part of $K_{a', \hat{b} \hat{c}}$ is non-zero
since $\epsilon_{1a'}$ and $\epsilon_{2a'}$ are non-zero.
Therefore we find that $\Omega_{mn\hat{c}} = 0$ from \eqref{eq:torsion11} 
and 
\eqref{eq:torsion12} is satisfied automatically for any
$K_{\hat{a},\hat{b} \hat{c}}$.
Using \eqref{eq:half_scalar}, the anti-self-dual part of $K_{\hat{a},
\hat{b} \hat{c}}$ becomes zero. 
Since the $U(1)_{L'} \times U(1)_{R'}$ $K_{a',\hat{b} \hat{c}}$ 
charge of $T_{\hat{a} \hat{b}} {}^{\hat{c}}$ is non-zero, 
the condition \eqref{eq:U1_charge_condition} implies 
$T_{\hat{a} \hat{b}} {}^{\hat{c}} = 0$. 
Then, using the relation \eqref{contorsion} the self-dual part of 
$K_{\hat{a},\hat{b} \hat{c}}$ is zero. 
The conditions \eqref{eq:torsion33} and \eqref{eq:torsion41} are satisfied 
automatically. 
The self-dual part of $K_{a',\hat{b}\hat{c}}$ belongs to $U(1)_{L'}$. 
In summary, we have the following conditions on the $\Omega$-background 
parameters $\Omega_{mna}$ and the contorsion $K_{a,bc}$ for the scalar 
supersymmetry generated by $\bar{Q}$: 
\begin{eqnarray}
\begin{aligned}
 & \delta^{\dot{\alpha}} {}_{B'} K_{a',\hat{b}
 \hat{c}} (\bar{\Sigma}^{\hat{b} \hat{c}})_{A'} {}^{B'} = -
\delta^{\dot{\beta}} {}_{A'}
\Omega_{mn a'} (\bar{\sigma}^{mn})^{\dot{\alpha}} {}_{\dot{\beta}}, \\
 & 
K_{a', \hat{b} \hat{c}} (\Sigma^{\hat{b} \hat{c}})^{\hat{A}} 
{}_{\hat{B}}
= - 4 i M_{a'} {}^{\hat{A}} {}_{\hat{B}}, \quad 
M_{a'} {}^{\hat{A}} {}_{\hat{B}} = 
\left(
\begin{array}{cc}
m_{a'} & 0 \\
0 & - m_{a'}
\end{array}
\right)
, \\
& \Omega_{mn \hat{a}} = K_{\hat{a}, \hat{b} \hat{c}} = K_{a, b'c'}
 = K_{a, \hat{b} c'} = 0, 
\end{aligned}
\label{eq:ht_sSUSY}
\end{eqnarray}
where $m_{a'}$ are real parameters. 
The theory has $\mathcal{N} = (0,2)$ supersymmetry\footnote{The notation 
$\mathcal{N} = (m,n)$ means that the theory has $m$ chiral, $n$
anti-chiral supercharges.} under the conditions \eqref{eq:ht_sSUSY}.
The explicit form of the supersymmetry transformation of fields are obtained 
by substituting \eqref{eq:ht_sSUSY} into \eqref{susy_transf2}. 
The result coincides with the transformation obtained in \cite{Ito:2011cr}. 

The R-symmetry Wilson line $(\mathcal{A}_{a'})^{\hat{A}}{}_{\hat{B}}$ 
and the contorsion $K_{a', \hat{b} \hat{c}}$ are 
related by \eqref{eq:Contorsion-WL}. 
The conditions on the contorsion in \eqref{eq:ht_sSUSY} are rewritten as
\begin{eqnarray}
\begin{aligned}
 & 
4 i \delta^{\dot{\alpha}} {}_{B'} (\mathcal{A})_{A'} {}^{B'} =
\delta^{\dot{\beta}} {}_{A'} \Omega_{mn} (\bar{\sigma}^{mn})^{\dot{\alpha}} {}_{\dot{\beta}}, 
\quad 
4 i \delta^{\dot{\alpha}} {}_{B'} (\bar{\mathcal{A}})_{A'} {}^{B'} =
\delta^{\dot{\beta}} {}_{A'} \bar{\Omega}_{mn} (\bar{\sigma}^{mn})^{\dot{\alpha}}
 {}_{\dot{\beta}}, \\
 & (\mathcal{A})^{\hat{A}} {}_{\hat{B}} 
= 
\left(
\begin{array}{cc}
m & 0 \\
0 & -m
\end{array}
\right), 
\quad 
(\bar{\mathcal{A}})^{\hat{A}} {}_{\hat{B}} 
= 
\left(
\begin{array}{cc}
\bar{m} & 0 \\
0 & - \bar{m}
\end{array}
\right), 
\label{eq:N2hypermass} \\
 & \Omega_{mn \hat{a}} = 
(\mathcal{A}_{\hat{a}})^A {}_B (\Sigma_{\hat{b} \hat{c}})^B {}_A 
= (\mathcal{A}_{a})^A {}_B (\Sigma_{b' c'})^B {}_A  
= (\mathcal{A}_a)^A {}_B (\Sigma_{\hat{b}c'})^B {}_A = 0, 
\end{aligned}
\end{eqnarray}
where 
$
\mathcal{A} = \frac{1}{\sqrt{2}} (\mathcal{A}_5 - i \mathcal{A}_6)
$
,
$
\bar{\mathcal{A}} = \frac{1}{\sqrt{2}} (\mathcal{A}_5 + i
\mathcal{A}_6)
$
and $m, \bar{m}, \Omega_{mn}, \bar{\Omega}_{mn}$ are defined similarly.
$(\mathcal{A})^{\hat{A}} {}_{\hat{B}}, (\bar{\mathcal{A}})^{\hat{A}} {}_{\hat{B}}$ 
are identified with the mass matrices of the hypermultiplet in the $\mathcal{N} = 2^{*}$
theory \cite{Nekrasov:2003rj, Ito:2011cr}.
The mass of the hypermultiplet is $\sqrt{m\bar{m}}$.
Mass perturbations in twisted $\mathcal{N} = 4$ theory are discussed in \cite{Labastida:1997xk}.

\subsubsection{Vafa-Witten twist}
We next consider the Vafa-Witten twist. 
The supercharges $Q^{A'}_{\alpha}, \bar{Q}^{\dot{\alpha}}_{A'}$ 
and the parameters $\zeta^{A'}_{\alpha},
\bar{\zeta}^{\dot{\alpha}}_{A'}$ are decomposed as in the case of the
half twist.
The supercharges $Q^{\hat{A}}_{\alpha},
\bar{Q}^{\dot{\alpha}}_{\hat{A}}$ are decomposed into 
$\bar{Q}'$, $\bar{Q}'_{mn}$ and $\bar{Q}'_{mn}$.
The parameters $\zeta^{\hat{A}}_{\alpha},
\bar{\zeta}^{\dot{\alpha}}_{\hat{A}}$ are decomposed in the same way.
The parallel spinor conditions for $\bar{\zeta}$, $\bar{\zeta}_{mn}$ and
$\zeta_m$ are \eqref{eq:ht_scalar_1}--\eqref{eq:ht_vector}. 
The conditions for $\bar{\zeta}', \bar{\zeta}'_{mn}$ and $\zeta'_m$ are 
\begin{eqnarray}
& & 
\left[
\delta^{\dot{\beta}} {}_{\hat{A}}
\Omega_{mna} (\bar{\sigma}^{mn})^{\dot{\alpha}} {}_{\dot{\beta}}
+ 
\delta^{\dot{\alpha}} {}_{\hat{B}} K_{a,\hat{b}
 \hat{c}} (\bar{\Sigma}^{\hat{b} \hat{c}})_{\hat{A}} {}^{\hat{B}} 
\right] \bar{\zeta}' = 0, \\
& & 
\left[
\delta^{\dot{\gamma}} {}_{\hat{A}}
\Omega_{mna} (\bar{\sigma}^{mn})^{\dot{\alpha}} {}_{\dot{\beta}}
(\bar{\sigma}^{pq})^{\dot{\beta}} {}_{\dot{\gamma}} + 
\delta^{\dot{\beta}} {}_{\hat{B}}
K_{a,\hat{b} \hat{c}}
(\bar{\Sigma}^{\hat{b} \hat{c}})_{\hat{A}} {}^{\hat{B}} 
(\bar{\sigma}^{pq})^{\dot{\alpha}} {}_{\dot{\beta}}
\right] \bar{\zeta}'_{pq} = 0, \\
\label{eq:vw-tensor}
& & 
\left[
\delta^{\dot{\gamma}} {}_{\hat{C}}
\Omega_{mna} (\sigma^{mn})_{\alpha} {}^{\beta} (\sigma^p)_{\beta \dot{\gamma}}
\epsilon^{\hat{A} \hat{C}} + 
\delta^{\dot{\beta}} {}_{\hat{C}} 
K_{a,\hat{b} \hat{c}} (\sigma^p)_{\alpha \dot{\beta}}
\epsilon^{\hat{B} \hat{C}} (\Sigma^{\hat{b} \hat{c}})^{\hat{A}} {}_{\hat{B}} 
\right] \zeta'_p = 0.
\label{eq:vw-vector}
\end{eqnarray}
The condition on the torsion for the scalar supersymmetry $\bar{\zeta}'$ is 
\begin{eqnarray}
 \delta^{\dot{\alpha}} {}_{\hat{B}} K_{a,\hat{b}
 \hat{c}} (\bar{\Sigma}^{\hat{b} \hat{c}})_{\hat{A}} {}^{\hat{B}} = -
\delta^{\dot{\beta}} {}_{\hat{A}}
\Omega_{mna} (\bar{\sigma}^{mn})^{\dot{\alpha}} {}_{\dot{\beta}}.
\label{eq:Vafa-Witten_scalar}
\end{eqnarray}
The condition \eqref{eq:Vafa-Witten_scalar} together with
\eqref{eq:half_scalar} implies that the rank of the contorsion 
matrices $K_{a' \hat{b} \hat{c}}$ reduces by two 
and $K_{a' \hat{b} \hat{c}}$ can be of the form  
\begin{eqnarray}
K_{a', \hat{b} \hat{c}} 
= 
\left(
\begin{array}{cccc}
0 & - k_{a'} & 0 & 0 \\
 k_{a'} & 0 & 0 & 0 \\
0 & 0 & 0 & 0 \\
0 & 0 & 0 & 0 
\end{array}
\right),
\end{eqnarray}
where $k_{a'}$ are non-zero parameters.
Therefore, two matrices within $\Omega_{mn\hat{a}} \ (\hat{a} = 7,8,9,10)$ remain non-zero
from the condition \eqref{eq:torsion11}.
From the representation of the $\Sigma$-matrices in the appendix, we can
take $\Omega_{mn9}$, $\Omega_{mn10}$ as non-zero matrices and
$\Omega_{mn7}$ and $\Omega_{mn8}$ as zero.
Then from \eqref{eq:half_scalar} and \eqref{eq:Vafa-Witten_scalar} we
get $K_{\hat{a}, \hat{b} \hat{c}} = 0$ except $K_{\hat{a}, 78} \
(\hat{a} = 9,10)$ 
and the condition \eqref{eq:torsion12} is satisfied automatically. 
The condition \eqref{eq:contorsion_commutation} holds since the contorsion 
and the $\Omega$-background matrices are identified by the relations 
\eqref{eq:Vafa-Witten_scalar} and \eqref{eq:half_scalar}, 
and the matrices $\Omega_{mna}$ commute with each other.
We find that the condition \eqref{eq:U1_charge_condition} is
satisfied and the \eqref{eq:torsion33}
reduces to the commutative relation of $\Omega_{mn7}$ and $\Omega_{mn8}$. 
The last condition \eqref{eq:torsion41} holds 
when the conditions \eqref{eq:torsion11}, \eqref{eq:torsion12},
\eqref{eq:contorsion_commutation} and \eqref{eq:U1_charge_condition} 
are satisfied.

We obtain the following conditions on the $\Omega$-background
parameters $\Omega_{mna}$ and the contorsion $K_{a,bc}$ for the scalar
supersymmetries generated by $\bar{Q}, \bar{Q}'$:
\begin{eqnarray}
\begin{aligned}
 & \delta^{\dot{\alpha}} {}_{B'} K_{a',\hat{b}
 \hat{c}} (\bar{\Sigma}^{\hat{b} \hat{c}})_{A'} {}^{B'} = -
\delta^{\dot{\beta}} {}_{A'}
\Omega_{mna'} (\bar{\sigma}^{mn})^{\dot{\alpha}} {}_{\dot{\beta}}, \\
 & \delta^{\dot{\alpha}} {}_{\hat{B}} K_{a',\hat{b}
 \hat{c}} (\bar{\Sigma}^{\hat{b} \hat{c}})_{\hat{A}} {}^{\hat{B}} = -
\delta^{\dot{\beta}} {}_{\hat{A}}
\Omega_{mna'} (\bar{\sigma}^{mn})^{\dot{\alpha}} {}_{\dot{\beta}}, 
\quad (a' = 5,6)
\\
 & \delta^{\dot{\alpha}} {}_{B'} K_{\hat{a},\hat{b}
 \hat{c}} (\bar{\Sigma}^{\hat{b} \hat{c}})_{A'} {}^{B'} = -
\delta^{\dot{\beta}} {}_{A'}
\Omega_{mn\hat{a}} (\bar{\sigma}^{mn})^{\dot{\alpha}} {}_{\dot{\beta}}, \\
 & \delta^{\dot{\alpha}} {}_{\hat{B}} K_{\hat{a},\hat{b}
 \hat{c}} (\bar{\Sigma}^{\hat{b} \hat{c}})_{\hat{A}} {}^{\hat{B}} = -
\delta^{\dot{\beta}} {}_{\hat{A}}
\Omega_{mn\hat{a}} (\bar{\sigma}^{mn})^{\dot{\alpha}} {}_{\dot{\beta}},
 \quad 
(\hat{a} = 9,10)
\\
& 
\Omega_{mn7} = \Omega_{mn8} 
 = K_{7, \hat{b} \hat{c}} = K_{8, \hat{b} \hat{c}}
= K_{a, b'c'} = K_{a, \hat{b} c'} 
= 0.
\end{aligned}
\label{eq:vw_sSUSY}
\end{eqnarray}
In terms of the R-symmetry Wilson line, these become
\begin{eqnarray}
\begin{aligned}
 & 4 i \delta^{\dot{\alpha}} {}_{B'} (\mathcal{A}_{a'})_{A'} {}^{B'} = -
\delta^{\dot{\beta}} {}_{A'}
\Omega_{mna'} (\bar{\sigma}^{mn})^{\dot{\alpha}} {}_{\dot{\beta}}, \\
 & 4 i \delta^{\dot{\alpha}} {}_{\hat{B}} (\mathcal{A}_{\hat{a}})_{\hat{A}} {}^{\hat{B}} = -
\delta^{\dot{\beta}} {}_{\hat{A}}
\Omega_{mna'} (\bar{\sigma}^{mn})^{\dot{\alpha}} {}_{\dot{\beta}}, 
\quad (a' = 5,6)
\\
 & 4 i \delta^{\dot{\alpha}} {}_{B'} (\mathcal{A}_{a'})_{A'} {}^{B'} = -
\delta^{\dot{\beta}} {}_{A'}
\Omega_{mn\hat{a}} (\bar{\sigma}^{mn})^{\dot{\alpha}} {}_{\dot{\beta}}, \\
 & 4 i \delta^{\dot{\alpha}} {}_{\hat{B}} (\mathcal{A}_{\hat{a}})_{\hat{A}} {}^{\hat{B}} = -
\delta^{\dot{\beta}} {}_{\hat{A}}
\Omega_{mn\hat{a}} (\bar{\sigma}^{mn})^{\dot{\alpha}} {}_{\dot{\beta}},
 \quad (\hat{a} = 9,10), \\
 & 
 (\mathcal{A}_7)^A {}_B (\Sigma_{\hat{b} \hat{c}})^B {}_A
= (\mathcal{A}_8)^A {}_B (\Sigma_{\hat{b} \hat{c}})^B {}_A
= (\mathcal{A}_a)^A {}_B (\Sigma_{b'c'})^B {}_A
= (\mathcal{A}_a)^A {}_B (\Sigma_{\hat{b}c'})^B {}_A 
= 0, \\
 & \Omega_{mn7} = \Omega_{mn8} = 0.
\end{aligned}
\label{eq:vw_sSUSY2}
\end{eqnarray}
As in the case of the half twist, two components of the tensor
supersymmetries are preserved when the conditions 
\eqref{eq:vw_sSUSY}, \eqref{eq:vw_sSUSY2} are satisfied.
Therefore the theory has $\mathcal{N} = (0,4)$ supersymmetry.

\subsubsection{Marcus twist}
Finally, we consider the Marcus twist. 
The supercharges $Q^{A'}_{\alpha}, \bar{Q}^{\dot{\alpha}} {}_{A'}$ 
and $Q^{\hat{A}}_{\alpha}, \bar{Q}^{\dot{\alpha}}_{\hat{A}}$ 
are decomposed into $Q, \bar{Q}$, $Q_{mn}, \bar{Q}_{mn}$ and $Q_m,
\bar{Q}_m$.
The parameters $\zeta^{A'}_{\alpha}, \bar{\zeta}^{\dot{\alpha}}_{A'}$
and $\zeta^{\hat{A}}_{\alpha}, \bar{\zeta}^{\dot{\alpha}}_{\hat{A}}$ are
decomposed similarly.
The parallel spinor conditions for $\bar{\zeta}$, $\bar{\zeta}_{mn}$ and
$\zeta_m$ are \eqref{eq:ht_scalar_1}--\eqref{eq:ht_vector}. 
The condition \eqref{parallel_spinor_omega} for 
$\zeta, \zeta_{pq}$ and $\bar{\zeta}_p$ are 
\begin{eqnarray}
& & 
\left[
\delta_{\beta} {}^{\hat{A}} \Omega_{mna} (\sigma^{mn})_{\alpha}
 {}^{\beta} + 
\delta_{\alpha} {}^{\hat{B}} K_{a, \hat{b} \hat{c}}
 (\Sigma^{\hat{b} \hat{c}})^{\hat{A}} {}_{\hat{B}}
\right] \zeta = 0, 
\\
& & 
\left[
\delta_{\alpha} {}^{\hat{A}}
\Omega_{mna} (\sigma^{mn})_{\alpha} {}^{\beta} (\sigma^{pq})_{\beta}
{}^{\alpha} + 
\delta_{\beta} {}^{\hat{B}}
K_{a,\hat{b} \hat{c}} (\sigma^{pq})_{\alpha} {}^{\beta}
(\Sigma^{\hat{b} \hat{c}})^{\hat{A}} {}_{\hat{B}} 
\right] \zeta_{pq} = 0, 
\label{eq:mt_tensor}
\\
& & 
\left[
\delta_{\gamma} {}^{\hat{C}}
\Omega_{mna} (\bar{\sigma}^{mn})^{\dot{\alpha}} {}_{\dot{\beta}}
 (\sigma^p)^{\dot{\beta} \gamma} \epsilon_{\hat{A} \hat{C}} + 
\delta_{\beta} {}^{\hat{C}}
K_{a, \hat{b} \hat{c}}
 (\bar{\Sigma}^{\hat{b} \hat{c}})_{\hat{A}} {}^{\hat{B}} (\sigma^p)^{\dot{\alpha}
 \beta} \epsilon_{\hat{B} \hat{C}} 
\right] \bar{\zeta}_p = 0.
\label{eq:mt_vector}
\end{eqnarray}
We examine the conditions on the two scalar supersymmetries generated by
$Q, \bar{Q}$.
The condition on the torsion for the parallel spinor $\zeta$ is 
\begin{eqnarray}
\delta_{\alpha} {}^{\hat{B}} K_{a, \hat{b} \hat{c}}
 (\Sigma^{\hat{b} \hat{c}})^{\hat{A}} {}_{\hat{B}} = 
- \delta_{\beta} {}^{\hat{A}} \Omega_{mna} (\sigma^{mn})_{\alpha}
 {}^{\beta}.
\label{eq:Marcus_scalar}
\end{eqnarray}
Using the relation \eqref{eq:Marcus_scalar} together with
\eqref{eq:half_scalar}, the condition \eqref{eq:torsion11} implies that 
the matrices $\Omega_{mn\hat{c}}$ vanish.
Then the condition \eqref{eq:torsion12} holds automatically.
The condition \eqref{eq:contorsion_commutation} is satisfied 
by using the relation \eqref{eq:Marcus_scalar}. 
Similarly $K_{\hat{a}, \hat{b} \hat{c}}$ is shown to be zero from 
$\Omega_{mn\hat{c}} = 0$. 
Then the conditions \eqref{eq:U1_charge_condition}, 
\eqref{eq:contorsion_commutation} and \eqref{eq:torsion41} are satisfied.

We get the following conditions on the $\Omega$-background
parameters $\Omega_{mna}$ and the contorsion $K_{a,bc}$ for the scalar
supersymmetries generated by $Q, \bar{Q}$:
\begin{eqnarray}
\begin{aligned}
 & \delta^{\dot{\alpha}} {}_{B'} K_{a',\hat{b}
 \hat{c}} (\bar{\Sigma}^{\hat{b} \hat{c}})_{A'} {}^{B'} = -
\delta^{\dot{\beta}} {}_{A'}
\Omega_{mna'} (\bar{\sigma}^{mn})^{\dot{\alpha}} {}_{\dot{\beta}}, \\
 & \delta_{\alpha} {}^{\hat{B}} K_{a', \hat{b} \hat{c}}
 (\Sigma^{\hat{b} \hat{c}})^{\hat{A}} {}_{\hat{B}}
= - \delta_{\beta} {}^{\hat{A}} \Omega_{mna'} (\sigma^{mn})_{\alpha}
 {}^{\beta}, \\
 & \Omega_{mn\hat{a}} = K_{\hat{a}, \hat{b} \hat{c}} = K_{a,b' c'} =
 K_{a, \hat{b} c'} = 0. \\
\end{aligned}
\label{eq:mt_sSUSY}
\end{eqnarray}
In terms of the R-symmetry Wilson line, the conditions become
\begin{eqnarray}
\begin{aligned}
 & 4 i \delta^{\dot{\alpha}} {}_{B'} (\mathcal{A}_{a'})_{A'} {}^{B'} = -
\delta^{\dot{\beta}} {}_{A'}
\Omega_{mna'} (\bar{\sigma}^{mn})^{\dot{\alpha}} {}_{\dot{\beta}}, \\
 & 4 i \delta_{\alpha} {}^{\hat{B}} (\mathcal{A}_{a'})^{\hat{A}} {}_{\hat{B}}
= - \delta_{\beta} {}^{\hat{A}} \Omega_{mna'} (\sigma^{mn})_{\alpha}
 {}^{\beta}, \\
 & \Omega_{mn\hat{a}} = 
(\mathcal{A}_{\hat{a}})^A {}_B (\Sigma_{\hat{b} \hat{c}})^B {}_A 
= (\mathcal{A}_a)^A {}_B (\Sigma_{b'c'})^B {}_A
= (\mathcal{A}_a)^A {}_B (\Sigma_{\hat{b}c'})^B {}_A = 0. \\
\end{aligned}
\label{eq:mt_sSUSY2}
\end{eqnarray}
In addition to the scalar supersymmetries, 
two components of the tensor supersymmetries are preserved when the 
conditions \eqref{eq:mt_sSUSY}, \eqref{eq:mt_sSUSY2} are satisfied.
Therefore the theory has $\mathcal{N} = (2,2)$ supersymmetry.

\subsection{Nekrasov-Shatashvili limit}
We have examined the scalar supersymmetries of the $\mathcal{N} = 4$
super Yang-Mills theory in the $\Omega$-background with the torsion. 
In this subsection, we study supersymmetries of the theory in the 
Nekrasov-Shatashvili limit of the $\Omega$-background \cite{Nekrasov:2009rc}. 
It is defined by the limit where
$\epsilon_{2a}$ (or $\epsilon_{1a}$)${}\to 0$ and keeping $\epsilon_{1a}$ 
(or $\epsilon_{2a}$) finite. 
In this limit, the super Poincar\'e symmetry of the two-dimensional 
subspace in four-dimensional spacetime is recovered.
We will study how supersymmetry is enhanced in each topological twist.

\paragraph{Half twist}
First, we consider the half twist where 
the contorsion and the $\Omega$-background matrices are related by \eqref{eq:half_scalar}.
We examine the parallel spinor conditions for the tensor and the vector supersymmetries
\eqref{eq:ht_tensor}, \eqref{eq:ht_vector}.

For the vector supersymmetry, 
by eliminating the contorsion in \eqref{eq:ht_vector},
we get 
\begin{eqnarray}
\left[
\Omega_{mna} (\sigma^{mn})_{\alpha} {}^{\beta} (\sigma^p)_{\beta \dot{\beta}}
 \varepsilon^{\dot{\beta} A'} + (\sigma^p)_{\alpha \dot{\alpha}}
 \varepsilon^{\dot{\alpha} B'}
 \Omega_{mna} (\bar{\sigma}^{mn})_{B'} {}^{A'} 
\right] \zeta_p = 0.
\label{eq:ht_scalar-vector}
\end{eqnarray}
This equation \eqref{eq:ht_scalar-vector} is written in terms of
$\epsilon_{1a}, \epsilon_{2a}$ as 
\begin{eqnarray}
 \epsilon_{1a} (\sigma^4 \zeta_1 - \sigma^3 \zeta_2)
+
 \epsilon_{2a} (\sigma^2 \zeta_3 - \sigma^1 \zeta_4) = 0.
\label{eq:ht_NS_vector}
\end{eqnarray}
In the Nekrasov-Shatashvili limit, $\epsilon_{2a} \to 0 \ (a = 5,6)$, 
the parameters $\zeta_3, \zeta_4$ satisfy the parallel spinor condition.
In the limit $\epsilon_{1a} \to 0 \ (a = 5,6)$, the parameters $\zeta_{1}, \zeta_2$
satisfy the condition.

We next consider parallel spinor condition for $\bar{\zeta}_{pq}$.
Using the relation \eqref{eq:half_scalar}, the condition is rewritten as 
\begin{equation}
\left[
\Omega_{mna} (\bar{\sigma}^{mn})^{\dot{\alpha}} {}_{\dot{\beta}}
(\bar{\sigma}^{pq})^{\dot{\beta}} {}_{A'} -
(\bar{\sigma}^{pq})^{\dot{\alpha}} {}_{\dot{\beta}} \Omega_{mna}
(\bar{\sigma}^{mn})^{\dot{\beta}} {}_{A'}
\right] \bar{\zeta}_{pq} = 0.
\label{eq:ht_scalar-tensor}
\end{equation}
This is the commutative relation between the matrices $\Omega_{mna}
(\bar{\sigma}^{mn})$ and $\bar{\zeta}_{pq} (\bar{\sigma}^{pq})$,
\begin{eqnarray}
[\Omega_{mna} \bar{\sigma}^{mn}, \bar{\zeta}_{pq} \bar{\sigma}^{pq}] =
 0.
\label{eq:ht_NS_tensor}
\end{eqnarray}
Since we have $\Omega_{mna} (\bar{\sigma}^{mn}) = i (\epsilon_{1a} +
\epsilon_{2a}) \tau_3$, the parameter $\bar{\zeta}_{12}$ 
satisfies \eqref{eq:ht_NS_tensor} as mentioned in section
\ref{half_twist} and $\bar{\zeta}_{13} = \bar{\zeta}_{14} = 0$.
Therefore $\mathcal{N} = (2,2)$ 
supersymmetry
 is preserved in the Nekrasov-Shatashvili limit.
The conserved supercharges of the $\mathcal{N} = (2,2)$ supersymmetry in the
half twist case is summarized in table \ref{tb:half}.
\begin{table}[t]
\begin{center}
\begin{tabular}{|l||c|c|c|}
\hline
Supercharge & Scalar & Tensor & Vector \\
\hline \hline
$\epsilon_{1a} \to 0 \ (a = 5,6)$ & $\bar{Q}$ & $\bar{Q}_{12}$ & $Q_1, Q_2$ \\
\hline
$\epsilon_{2a} \to 0 \ (a = 5,6)$ & $\bar{Q}$ & $\bar{Q}_{12}$ & $Q_3, Q_4$ \\
\hline
\end{tabular}
\caption{Conserved supercharges in the half twist.}
\label{tb:half}
\end{center}
\end{table}

\paragraph{Vafa-Witten twist}.
The parallel spinor conditions for $\bar{\zeta}'_{mn}$ and $\zeta'_{m}$ are 
the same as those for $\bar{\zeta}_{mn}$ and $\zeta_{m}$ in the half twist. 
Therefore $\mathcal{N} = (4,4)$ supersymmetry is preserved in 
the Nekrasov-Shatashvili limit.
The conserved supercharges are found in table \ref{tb:vw}.
\begin{table}[t]
\begin{center}
\begin{tabular}{|l||c|c|c|}
\hline
Supercharge & Scalar & Tensor & Vector \\
\hline \hline
$\epsilon_{1a} \to 0 \ (a=5,6,7,8)$ & $\bar{Q}, \bar{Q}'$  & $\bar{Q}_{12},
		 \bar{Q}'_{12}$ & $Q_1, Q_2, Q_1', Q_2'$ \\
\hline
$\epsilon_{2a} \to 0 \ (a=5,6,7,8)$ & $\bar{Q}, \bar{Q}'$ & $\bar{Q}_{12},
		 \bar{Q}'_{12}$ & $Q_3, Q_4, Q_3', Q_4'$ \\
\hline
\end{tabular}
\caption{Conserved supercharges in the Vafa-Witten twist.}
\label{tb:vw}
\end{center}
\end{table}

\paragraph{Marcus twist}
The parallel spinor conditions for $\bar{\zeta}_{pq}$, $\zeta_{p}$ 
have been written down. 
The parallel spinor condition for $\zeta_{pq}$ is 
obtained by replacing $\bar{\sigma}_{mn}$ with $\sigma_{mn}$ in 
\eqref{eq:ht_NS_tensor}. Then $\zeta_{12}$ satisfies the equation 
\eqref{eq:ht_NS_tensor} and $\zeta_{13} = \zeta_{14} = 0$. 
The condition for $\bar{\zeta}_p$ is 
\begin{equation}
\epsilon_{1a} (\sigma^2 \bar{\zeta}_3 - \sigma^1 \bar{\zeta}_4) 
+ 
\epsilon_{2a} (\sigma^4 \bar{\zeta}_1 - \sigma^3 \bar{\zeta}_2)
= 0.
\end{equation}
Therefore $\mathcal{N} = (4,4)$ supersymmetry is preserved in the
Nekrasov-Shatashvili limit. 
The conserved supercharges are summarized in table \ref{tb:marcus}.
\begin{table}[t]
\begin{center}
\begin{tabular}{|l||c|c|c|}
\hline
Supercharge & Scalar & Tensor & Vector \\
\hline \hline
$\epsilon_{1a} \to 0 \ (a=5,6)$ & $\bar{Q}, Q$  & $\bar{Q}_{12},
		 Q_{12}$ & $Q_1, Q_2, \bar{Q}_1, \bar{Q}_2$ \\
\hline
$\epsilon_{2a} \to 0 \ (a=5,6)$ & $\bar{Q}, Q$ & $\bar{Q}_{12},
		 Q_{12}$ & $Q_3, Q_4, \bar{Q}_3, \bar{Q}_4$ \\
\hline
\end{tabular}
\caption{Conserved supercharges in the Marcus twist.}
\label{tb:marcus}
\end{center}
\end{table}

\section{Conclusions and Discussion}
In this paper we studied ten-dimensional $\mathcal{N}=1$ super 
Yang-Mills theory in the curved background with the torsion. 
We investigated the dimensional reduction to four dimensions where 
the torsion is introduced along the internal space such that four-dimensional 
gauge invariance is preserved. 
Requiring supersymmetry, it has been shown that the torsion 
obeys the constraints and also modifies the parallel spinor conditions.
In particular we have studied the torsion in the ten-dimensional
$\Omega$-background, where we can identify the R-symmetry Wilson line 
gauge field with the torsion.
We solved the modified parallel spinor conditions for the topological twists 
(the half-twist, the Vafa-Witten twist and the Marcus twist) of
$\mathcal{N}=4$ supersymmetry. We found the solutions of the deformation 
parameters of the $\Omega$-background and the Wilson line gauge fields. 
We obtained the preserved supersymmetries for these twists. 

In order to construct the deformed topological field theory associated 
with the twists, 
it is necessary to extend the deformed scalar supersymmetry to 
the off-shell supersymmetry. 
In a subsequent paper \cite{ItNaSa2}, we will discuss the off-shell 
structure of the twisted deformed theories and their instanton 
effective action. 

One can consider the curved background admitting 
the (conformal) Killing spinor 
associated with 
super(conformal)symmetry transformations.
In the case without the torsion, 
the parallel spinor conditions lead to 
the Ricci-flatness of the geometry. 
For the case with the Killing spinor, the associated geometries 
have been classified  
in \cite{Blau:2000xg}. 
The conformal Killing spinor in $\mathrm{S}^{4}$ 
\cite{Pestun:2007rz,Okuda:2010ke} and other geometries 
\cite{Festuccia:2011ws,Dumitrescu:2012ha,Hama:2012bg}
are used to discuss the localization. 
It would be interesting to introduce the torsion for the geometry
admitting (conformal) Killing spinor conditions and examine the
deformed super(conformal)symmetry.

It would also  be an interesting problem to understand the ten-dimensional deformed theory
in superstring theory.
The stringy realization of the $\Omega$-background allows us to
analyze the various dimensional
system \cite{Jafferis:2007sg,Awata:2009dd,Ne-M,Billo:2009di,Fucito:2009rs,Billo:2010mg}.
{}From the dimensional reduction to
various dimensions, we can obtain the $\Omega$-deformed 
gauge theories other than four dimensions in a systematic way.
One can also study their Nekrasov-Shatashvili limit, where the various
$\Omega$-deformed BPS states exist in the deformed theories 
\cite{Ito:2011ta, Ito:2011wv,Hellerman:2012zf,Bulycheva:2012ct}. 
From the supersymmetry constructed in this work, one can compute the
central charges for the BPS states, 
which are important to understand
the integrability structure of the deformed theory.

\subsection*{Acknowledgements} We would like to thank T.~Saka for
collaboration in an early stage of the work.
The work of K.~I. is supported in part by Grant-in-Aid
for Scientific Research from the Japan Ministry of Education, Culture,
Sports, Science and Technology. 
The work of H.~N. is supported in part by National Science Council
and National Center for Theoretical Sciences, Taiwan, R.O.C.

\begin{appendix}
\def\thesection{Appendix}
\section{Dirac matrices in four and six dimensions} \label{sec:sigma}
\def\thesection{A}
In this appendix, 
we present our conventions of the Dirac matrices in four- and 
six-dimensional spaces with the Euclidean signature.
The Dirac matrices $\sigma^m_{\alpha\dot{\alpha}}$ and $\bar{\sigma}^{m \dot{\alpha}\alpha}$ 
 in four dimensions are defined 
 by 
\begin{align}
\sigma^{m}
&=
\big(i\tau^{1}, i\tau^{2}, i\tau^{3}, \boldsymbol{1}_{2}\bigr),
&
\bar{\sigma}^{m}
&=
\bigl(-i\tau^{1}, -i\tau^{2}, -i\tau^{3}, \boldsymbol{1}_{2}\bigr),
\end{align}
where $\tau^{i} $ ($i=1,2,3$) are the Pauli matrices
and $\boldsymbol{1}_2$ denotes the $2 \times 2$ identity matrix.
We define the Lorentz generators $\sigma^{mn}$ and $\bar{\sigma}^{mn}$ by 
\begin{align}
\sigma^{mn}
&=
\frac{1}{4}\bigl(\sigma^{m}\bar{\sigma}^{n}-\sigma^{n}\bar{\sigma}^{m}\bigr),
&
\bar{\sigma}^{mn}
&=
\frac{1}{4}\bigl(\bar{\sigma}^{m}\sigma^{n}-\bar{\sigma}^{n}\sigma^{m}\bigr).
\end{align}
The Dirac matrices $\Sigma^{aAB}$ and $\bar{\Sigma}^{a}_{AB}$ in six dimensions are defined by 
\begin{align}
\Sigma^{5}
&=
\begin{pmatrix} i\tau^{2} & 0 \\ 0 & i\tau^{2} \end{pmatrix},
&
\Sigma^{6}
&=
\begin{pmatrix} \tau^{2} & 0 \\ 0 & -\tau^{2} \end{pmatrix},
&
\Sigma^{7}
&=
\begin{pmatrix} 0 & -\tau^{3} \\ \tau^{3} & 0 \end{pmatrix},
\notag\\[2mm]
\Sigma^{8}
&=
\begin{pmatrix} 0 & i\boldsymbol{1}_{2} \\ -i\boldsymbol{1}_{2} & 0 
\end{pmatrix},
&
\Sigma^{9}
&=
\begin{pmatrix} 0 & - \tau^{1} \\ \tau^{1} & 0 \end{pmatrix},
&
\Sigma^{10}
&=
\begin{pmatrix} 0 & \tau^{2} \\ \tau^{2} & 0 \end{pmatrix},
\notag\\[2mm]
\bar{\Sigma}^{5}
&=
\begin{pmatrix} -i\tau^{2} & 0 \\ 0 & -i\tau^{2} \end{pmatrix},
&
\bar{\Sigma}^{6}
&=
\begin{pmatrix} \tau^{2} & 0 \\ 0 & -\tau^{2} \end{pmatrix},
&
\bar{\Sigma}^{7}
&=
\begin{pmatrix} 0 & \tau^{3} \\ -\tau^{3} & 0 \end{pmatrix},
\notag\\[2mm]
\bar{\Sigma}^{8}
&=
\begin{pmatrix} 0 & i\boldsymbol{1}_{2} \\ -i\boldsymbol{1}_{2} & 0 
\end{pmatrix},
&
\bar{\Sigma}^{9}
&=
\begin{pmatrix} 0 &  \tau^{1} \\ - \tau^{1} & 0 \end{pmatrix},
&
\bar{\Sigma}^{10}
&=
\begin{pmatrix} 0 & \tau^{2} \\ \tau^{2} & 0 \end{pmatrix},
\end{align}
The Lorentz generators $\Sigma^{ab}$ and $\bar{\Sigma}^{ab}$ are defined by 
\begin{align}
\Sigma^{ab}
&=
\frac{1}{4}\bigl(\Sigma^{a}\bar{\Sigma}^{b}-\Sigma^{b}\bar{\Sigma}^{a}\bigr),
&
\bar{\Sigma}^{ab}
&=
\frac{1}{4}\bigl(\bar{\Sigma}^{a}\Sigma^{b}-\bar{\Sigma}^{b}\Sigma^{a}\bigr).
\end{align}

\end{appendix}

\end{document}